\documentclass[12pt,epic,eepic,a4paper]{iopart}
\usepackage{amssymb}
\usepackage{amsfonts}

\jl{1}

\def\<{\langle}
\def\>{\rangle}

\def\Beta{{\rm B}}
\def\SG{{\mathfrak  S}}

\def\N{{\mathbb N}}

\def\Z{{\mathbb Z}}

\def\R{{\mathbb R}}

\def\pf{{\rm pf}}

\def\H{{\rm H}}
\def\If{{\rm I}}

\def\DET{{\rm Det}}

\begin{document}

\title[Hankel hyperdeterminants]{Hankel hyperdeterminants and Selberg integrals}
\author{Jean-Gabriel Luque and  Jean-Yves Thibon}

\address{Institut Gaspard Monge, Universit\'e de Marne-la-Vall\'ee,
 77454 Marne-la-Vall\'ee cedex 2, France}

\begin{abstract}
We investigate the simplest class of  hyperdeterminants defined by
Cayley in the case of Hankel hypermatrices (tensors of the form
$A_{i_1i_2\ldots i_k}=f(i_1+i_2+\cdots+i_k)$). It is found that
many classical properties of Hankel determinants can be
generalized, and a connection with Selberg type integrals is
established. In particular,  Selberg's original formula amounts to
the evaluation of all Hankel hyperdeterminants built from the
moments of the Jacobi polynomials. Many higher-dimensional
analogues of classical Hankel determinants are evaluated in closed
form. The Toeplitz case is also briefly discussed. In physical
terms, both cases are related to the partition functions of
one-dimensional Coulomb systems with logarithmic potential.
\end{abstract}

\parindent0mm

\section{Introduction}

Although determinants have been in use since the mid-eighteenth
century, it took almost one hundred years before the modern
notation as square arrays was introduced by Cayley \cite{Ca0}.
Then, it was not long before Cayley raised the question of
extending the notion of determinant to higher-dimensional arrays
(e.g., cubic matrices $A_{ijk}$), and proposed several answers,
under the name {\it hyperdeterminants} \cite{Ca1,Ca2}.

The most sophisticated notion of hyperdeterminant has been the
object of recent investigations, summarized in the book
\cite{GKZ}. However, the simplest possible generalization of the
determinant, defined for a $k$th order tensor on an
$n$-dimensional space by the $k$-tuple alternating sum (which
vanishes for odd $k$)
\begin{equation}\label{defDET}
\DET_k(A)=\frac1{n!}\sum_{\sigma_1,\cdots,\sigma_k\in \SG_n}
\epsilon(\sigma_1)\cdots\epsilon(\sigma_k) \prod_{i=1}^n
A_{\sigma_1(i) \cdots \sigma_k(i)}
\end{equation}
has been almost forgotten. After the book by Sokolov \cite{So1}
which contains an exhaustive list of references up to 1960, we
have found only \cite{So2,Hau,Bar,Gh}. These references contain
evaluations of a few higher dimensional analogues of some
classical determinants (Vandermonde, Smith, \dots). However, the
analogues of Hankel determinants do not seem to have been
investigated.

In this article, we shall compute the hyperdeterminantal analogues
of various classical Hankel determinants.
The elements of these determinants will in
general be  combinatorial numbers or orthogonal polynomials. Our
main technique will be the use of integral representations. We
shall see that the relevant tool is Selberg's integral and its
generalizations, mainly in the form given by Kaneko. More that
this, we can say that the knowledge embodied in Selberg's formula
and its limiting cases amounts  to a closed form evaluation of all
Hankel hyperdeterminants built from the moment sequences
associated with the classical orthogonal polynomials.

A more general class of hyperdeterminants is given by the
partition functions $Z_n(\beta)$ of log-potential Coulomb systems,
when $\beta$ is an even integer. When the particle-background
interaction does not lead to a Selberg integral, the partition
function can usually be evaluated in a more or less closed form only
for $\beta=1,2,4$ (as a Pfaffian or a determinant). The evaluation
of general hyperdeterminants  is of course much more difficult, but
their simple transformation properties leave some hope that at
least some of them may be evaluated by higher-dimensional
analogues of the algebraic techniques working for Hankel
determinants.

The consideration of the partition functions of similar Coulomb
systems with the particles confined on a circle suggests
immediately the following definition. A tensor $T_{i_1\ldots
i_k}^{j_1\ldots j_k}$ will be called a {\it Toeplitz tensor} if
\begin{equation}
T_{i_1\ldots i_k}^{j_1\ldots j_k} =f(i_1+\cdots+i_k - j_1-\cdots
-j_k)\,.
\end{equation}
Indeed, when $\beta$ is an even integer, the partition function
turns out to be a Toeplitz hyperdeterminant.

\section{Hankel hyperdeterminants}

Let $(A_{i_1\cdots i_k})_{0\leq i_1,\cdots,i_k\leq n-1}$ be a
tensor of order $k$ and dimension $n$. The tensor $A$ is said to
be a Hankel tensor if $A_{i_1 \cdots i_k}=f(i_1+\cdots+i_k)$.

Let us now fix some sequence $c=(c_n)_{n\ge 0}$, and consider the
hyperdeterminants
\begin{equation}
D_n^{(k)}(c)=\DET_{2k}(c_{i_1+\cdots+i_{2k}})_{0\le i_p \le
n-1}
\end{equation}
as defined by formula (\ref{defDET}), in which $\SG_n$ is the symmetric group
and $\epsilon(\sigma)$ the signature of a permutation $\sigma$.
For $k=1$, this is an ordinary Hankel determinant. For $n=2$, it
is easy to derive the expression
\begin{equation}\label{apolar}
D_2^{(k)}(c)= \frac12 \sum_{i=0}^{2k}(-1)^i\left(2k\atop
i\right)c_ic_{2k-i}
\end{equation}
whose right-hand side is  well-known  in classical invariant
theory (it is one-half of the apolar covariant of the binary form
$f(x,y)=\sum_i {2k\choose i}c_ix^i y^{2k-i}$ with itself, see
\cite{KR}).

The case $c_n=n!$ will be used as a running exemple throughout
this paper. Using (\ref{apolar}), we can give our first
illustration of a higher-order determinant
\begin{equation}
D_2^{(k)}(c)=\frac12 \sum_{i=0}^{2k}(-1)^i(2k)!=\frac12 (2k)!
\end{equation}
which will provide a check for the general case.

Let now $\mu$ be the linear functional on the space of polynomials
in one variable such that $\mu(x^n)=c_n$. We extend it to
polynomials in several variables by setting $\mu_n(x_1^{m_1}\cdots
x_n^{m_n}) =c_{m_1}\cdots c_{m_n}$. Then, using the expansion of
the Vandermonde determinant
\begin{equation}
\Delta(x)=\prod_{i>j}(x_i-x_j)=
\sum_{\sigma\in\SG_n}\epsilon(\sigma)
\sigma(x_n^{n-1}x_{n-1}^{n-2}\cdots x_{2})
\end{equation}
where a permutation $\sigma$ acts on a monomial by sending each
$x_i$ on $x_{\sigma(i)}$, it is easily seen that
\begin{equation}
D_n^{(k)}(c)={1\over n!}\mu_n(\Delta^{2k}(x))\,.
\end{equation}
Expanding each factor $(x_i-x_j)^{2k}$ by the binomial theorem, we
obtain
\begin{equation}
D_n^{(k)}(c)=\frac1{n!}\sum_{M=(m_{ij})} (-1)^{|M|} \prod_{i>j}
\left(2k\atop m_{ij}\right) \prod_{p=1}^nc_{\alpha_p(M)}
\end{equation}
where $M$ runs over all strictly lower triangular integer matrices
such that $0\le m_{ij} \le 2k$, $|M|=\sum_{i>j}m_{ij}$, and
\begin{equation}
\alpha_p(M)=2k(p-1)+\sum_{i=p+1}^{n}m_{ip} -\sum_{j=1}^{p-1}m_{pj}\,.
\end{equation}
This extends (\ref{apolar}) and provides a faster algorithm than
the definition.

Now, if $\mu$ is a measure on the real line, then
\begin{equation}\label{kHeine}
D_n^{(k)}(c)={1\over n!} \int_{\R^n} \Delta^{2k}(x)
d\mu(x_1)\cdots d\mu(x_n)\,.
\end{equation}
When $k=1$, this is a well-known formula due to Heine \cite{He}.
For arbitrary $k$, the integral can be evaluated in closed form in
many interesting cases by means of Selberg's integral formula
\cite{Sel} which gives, for
\begin{equation}
S_n(a,b,k)=\int_0^1\cdots \int_0^1 |\Delta(x)|^{2k}
\prod_{i=1}^nx_i^{a-1}(1-x_i)^{b-1} dx_i
\end{equation}
the value
\begin{equation}
S_n(a,b,k)=\prod_{j=0}^{n-1}
{\Gamma(a+jk)\Gamma(b+jk)\Gamma((j+1)k+1) \over
\Gamma(a+b+(n+j-1)k)\Gamma(k+1)}\,.
\end{equation}
The formula is valid, when defined,  for complex values of $k$ as
well, but it is interesting to observe that all its known proofs
start with the assumption that $k$ is a positive integer, and then
extend the result by means of Carlson's theorem (see, e.g.,
\cite{Meh1}).

Hence, Selberg's integral computes precisely the Hankel
hyperdeterminants $D_n^{(k)}(c)$ when the $c_n$ are the moments of
the measure $d\mu(x)={\bf 1}_{[0,1]}x^{a-1}(1-x)^{b-1}dx$ (the B\^eta distribution). 
The orthogonal polynomials for this
measure are, up to a simple change of variables,  the Jacobi
polynomials $P_n^{(a-1,b-1)}(1-2x)$. Hence, we can as well compute
$D_n^{(k)}(c)$ for the moments of the ordinary Jacobi polynomials,
and their limiting cases (Laguerre and Hermite). The appropriate
variants of Selberg's formula are listed in \cite{Meh1}, 17.6.

When the measure $d\mu(x)$ does not lead to a known variant of
Selberg's formula (such as Aomoto's and Kaneko's generalizations,
which are discussed below), it is sometimes possible to evaluate
the hyperdeterminant of order 4 ($k=2$) from the knowledge of the
scalar products $\<P_n,P'_m\>$ of the corresponding (monic)
orthogonal polynomials and their derivatives. Indeed, it is
classical (see \cite{Meh2}) that
\begin{equation}
\Delta(x)^4=\det(P_{i-1}(x_j)|P'_{i-1}(x_j))
\end{equation}
where, in the right-hand side, we mean the $2n\times 2n$-matrix
with $i=1,\ldots,2n$, $j=1,\ldots,n$, 
whose first $n$ columns are the $P_i$
and the last $n$ ones their derivatives. Using one of  de Bruijn's
formulae (see \cite{Meh1}, A.18.7), we can write the
hyperdeterminant as a Pfaffian
\begin{equation}\label{det4}
\DET_4(c_{i+j+k+l})|_{0}^{n-1}=\pf(M_{i j})|_{0}^{2n-1}
\end{equation}
where  $M$ is the skew symmetric matrix such that
\begin{equation}\label{matpprim}
M_{ij}=\langle P_i,P'_j\rangle=\int_a^bP_i(x)P'_j(x)d\mu(x)
\end{equation}
if $i<j$. For example, if $d\mu(x)=e^{-x}dx$ is the Laguerre
measure on $[0,\infty)$, whose monic orthogonal polynomials are
given by the generating series
\begin{equation}
\sum_{n\ge 0}\tilde L_n(x){(-t)^n\over n!}={e^{xt\over t-1}\over
1-t},
\end{equation}
an easy calculation gives
\begin{equation}
 \sum_{n,m\ge 0}\langle\tilde L_n(x),\tilde L'_m(x)\rangle{(-1)^{m+n}\over n!m!}t^ns^m
={s\over(1-s)(1-st)}
\end{equation}
which leads to
\begin{equation}
\DET_4((i+j+k+l)!)|_{0}^{n-1}=\pf
((-1)^{i+j-\delta_{i>j}}i!j!)|_{0}^{2n-1}
\end{equation}
 with
$\delta_{i>j}=1$ if $i>j$ and $0$ otherwise. The calculation of
the Pfaffian is straightforward, and we obtain
\begin{equation}
\DET_4((i+j+k+l)!)|_{0}^{n-1}=\prod_{i=0}^{2n-1}i!
\end{equation}
a special case of the general formula (\ref{factorielles}) derived
below from Selberg's integral.

For later reference, let us recall that the Hankel determinants
$D_n^{(1)}(c)$ are the products of the squared norms of the monic
orthogonal polynomials $P_n$, and that, more generally, the
shifted Hankel determinants $D_{n;r}^{(1)}(c)=D_n^{(1)}(c^{(r)})$,
associated to the shifted sequences $c^{(r)}_n=c_{n+r}$ are given
by
\begin{equation}\label{detshift}
D_{n;r}^{(1)}(c)=\det\left. \left(\<x^rP_i,P_j\>\right)
\right|_0^{n-1}\,.
\end{equation}
Similarly, the shifted hyperdeterminants of order $4$ can be
reduced to Pfaffians
\begin{equation}
D_{n;r}^{(2)}(c)=\pf\left.
\left((-1)^{\delta_{i<j}}\<x^rP_i,P'_j\>\right)
\right|_0^{2n-1}\,.
\end{equation}

Finally, let us remark that when the measure can be written in the
form $d\mu(x)=C e^{-\lambda V(x)}dx$, the integral (\ref{kHeine})
represents the partition function of a one-component log-potential
Coulomb system on the line, evaluated at $\beta=2k$ (see, e.g.,
\cite{Forr}). It is a common feature of most of these systems that
the partition function may be evaluated in closed form only for
$\beta=1,2,4$ (the case $\beta=1$ can be formulated in terms of
Pfaffians of bimoments of skew-orthogonal polynomials, see
\cite{Meh2}, and will not concern us here).

Similarly, the partition function of a one-component Coulomb
system of $n$ identical particles confined on the unit circle has
the general form
\begin{equation}
Z_n(\beta)=C_n{1\over (2\pi i)^n}\oint {dz_1\over z_1}\cdots\oint
{dz_n\over z_n} |\Delta(z)|^\beta \prod_{j=1}^N e^{-\beta
V(z_j)}\,.
\end{equation}
For $\beta=2k$, this is, up to a scalar factor,
 the hyperdeterminant of the Toeplitz
tensor associated to the bi-infinite sequence
\begin{equation}
c_n={1\over 2\pi i} \oint_{|z|=1}z^{-n}e^{-\beta V(z)} {dz\over z}
\end{equation}
As above, the knowledge of the appropriate orthogonal polynomials
allows one to evaluate the determinant $(k=1)$ and sometimes the
4-fold hyperdeterminant.

\section{Examples involving combinatorial numbers}

The evaluation of Hankel determinants built on classical sequences
of combinatorial numbers arises in many contexts (see
\cite{Kra,VD} and references therein). Also, recent work on the
theory of coherent states has led to the discovery of integral
representations of many such sequences, in a form directly
relevant to the evaluation of their Hankel hyperdeterminants
\cite{PSi,PSo}.

\subsection{Factorials and $\Gamma$-functions}

As a warm-up, let us start with the already considered sequence
\begin{equation}
c_n=n!=\Gamma(n+1)=\int_0^\infty x^n e^{-x}dx\,.
\end{equation}
Then, using the Laguerre-Selberg integral (see \cite{Meh1},
(17.6.5))
\begin{equation}\label{LagSel}
\fl LS_n(\alpha,\gamma)
=
\int_{(0,\infty)^n}|\Delta(x)|^{2\gamma}
\prod_{j=1}^nx_j^{\alpha-1}e^{-x_j} dx_j
=
\prod_{j=0}^{n-1}{\Gamma(1+\gamma+j\gamma)\Gamma(\alpha+j\gamma)
\over \Gamma(1+\gamma)}
\end{equation}
we find
\begin{equation}\label{factorielles}
{\rm F}(n,k):=D_n^{(k)}(c) ={1\over n! k!^n
}\prod_{j=0}^{n-1}(k+kj)!(kj)!
\end{equation}
thus recovering the classical evaluation
$D_n^{(1)}(c)=[1!2!\cdots(n-1)!]^2$ (see, e.g., \cite{VD})
 of the Hankel determinant. At no more cost, we can take
\begin{equation}
c_n=\Gamma(n+1+\alpha)
\end{equation}
corresponding to the measure $d\mu(x)=x^\alpha e^{-x}dx$ and
obtain
\begin{equation}\label{Gamma}
D_n^{(k)}(c) ={1\over n! k!^n }\prod_{j=0}^{n-1} {(k+jk)!\,
\Gamma(\alpha+1+jk)}
\end{equation}
which, for $\alpha=r$ a positive integer, gives the shifted
hyperdeterminant $D_{n;r}^{(k)}$ for factorials.

\subsection{Catalan numbers}

Here, we take  $c_n=C_n={1\over n+1}\left(2n\atop n\right)$.
Penson and Sixdeniers give in \cite{PSi} an integral
representation of these numbers
\begin{equation}
C_n=\frac1{2\pi}\int_0^4x^n\sqrt{4-x\over x}dx={2^{2n+1}\over
\pi}\Beta (n+\case12,\case32)\,.
\end{equation}
The Hankel hyperdeterminant can be written in the form
\begin{eqnarray}
D_n^{(k)}(c) &={2^{2kn(n-1)+n}\over
n!\pi^n}S_n(\case12,\case32,k)\nonumber
\\
&={2^{kn(n-1)-n}\over
n!k!^n}\prod_{j=0}^{n-1}{(k+kj)!(2kj+1)!!(2kj-1)!!\over(1+k(n+j-1))!}
\end{eqnarray}
where $(2n+1)!!=1\cdot 3\cdots (2n+1)$. The Hankel
hyperdeterminants of shifted Catalan numbers can be obtained
similarly. Indeed, replacing $c_n$ by $c_{r+n}$ leads to another
Selberg integral
\begin{equation}\label{shiftcat}
 D_{n;r}^{(k)}(c)= {2^{2kn(n-1)+n(2r+1)}\over
n!\pi^n}S_n(r+\case12,\case32,k)
\end{equation}
In the case $k=1$, the Hankel determinants of  shifted Catalan
numbers have been computed  by Desainte-Catherine and Viennot
\cite{DCV} with the aim to enumerate the Young tableaux whose
columns consists of an even number of elements and have height at
most $2n$. A natural question is  to find what is counted by
hyperdeterminants of shifted Catalan numbers.

\subsection{Central binomial coefficients}

In \cite{PSo}, Penson and Solomon give the representation
\begin{equation}
{2n\choose n}={1\over \pi}\int_0^4x^n [x(4-x)]^{-1/2}dx
={4^n\over\pi}\Beta(n+\case12,\case12)
\end{equation}
so that, for $c_n={2n\choose n}$,
\begin{equation}
D_n^{(k)}(c)={4^{kn(n-1)+nr}\over n!\pi^n}S_n(\case12,\case12,k)\,.
\end{equation}

Similarly, the shifted hyperdeterminants are given by
\begin{equation} D_{n;r}^{(k)}={4^{kn(n-1)+nr}
\over n!\pi^n}S_n(r+\case12,\case12,k)\,.
\end{equation}

\subsection{The sequence $(2n)!/n!$}

Here, we find in \cite{PSo} that
\begin{equation}
{(2n)!\over n!}= {1\over 2\sqrt{\pi}} \int_0^\infty x^n
e^{-x/4}x^{-1/2}dx\,.
\end{equation}
Setting $x=4y$ and using the Laguerre-Selberg integral
(\ref{LagSel}), we obtain the shifted hyperdeterminants as
\begin{eqnarray}
D_{n;r}^{(k)}(c) &=
\pi^{-\case{n}{2}}4^{n[k(n-1)+r]}LS_n(r+\case12,k)\nonumber\\ &={
2^{\frac32kn(n-1)+rn}  \over n!k!^n  }
\prod_{j=0}^{n-1}(k(1+j))!(2(kj+r)-1)!!\,.
\end{eqnarray}

\subsection{Bell numbers and polynomials}

We now take $c_n=b_n(a)$, the (one-variable) Bell polynomials,
defined by
\begin{equation}
b_0(a)=1 \mbox{ and }b_n(a)=\sum_{k=1}^nS(n,k)a^k
\end{equation}
where the $S(n,k)$ are the Stirling numbers of the second kind (so
that $b_n(1)$ are the Bell numbers). These are the moments of the
discrete measure
\begin{equation}
d\mu_a(x)=e^{-a}\sum_{k\ge 0}{a^k\over k!}\delta(x-k)
\end{equation}
for which the Charlier polynomials are the orthogonal system
(cf. \cite{KoSw}). The monic Charlier polynomials $C^{(a)}_n(x)$
satisfy
\begin{equation}
\<C_n^{(a)},C_n^{(a)}\>=a^n n!
\end{equation}
whence the classical evaluation of the Hankel determinants
\cite{Rad}
\begin{equation}
D_n^{(1)}=a^{n(n-1)/2}\prod_{j=0}^{n-1} j!\,.
\end{equation}
However, no analogue of Selberg's integral is known for the
measure $d\mu_a$. So, the best that we can do is to evaluate the
fourth-order hyperdeterminants by means of formula (\ref{det4}).
To this aim, we need the scalar products $\< C_n^{(a)},
C_m^{(a)}{}' \>$, which can be easily obtained from the generating
function
\begin{equation}
C(u,x;a)=\sum_{n\ge 0} C_n^{(a)}(x){u^n\over n!}
=e^{-au}(1+u)^x\,.
\end{equation}
Taking the scalar product of this expression with ${\partial
C(v,x;a)\over \partial x}$, we find that
\begin{equation}
\<C_n^{(a)},C_m^{(a)}{}'\>
=
\left\{
\begin{array}{cc}
(-1)^{m-n+1}{a^n m!\over m-n} & \mbox{\rm if $m>n$}\\ 0 &
\mbox{\rm otherwise}\,.
\end{array}
\right.
\end{equation}
It would remain to find a closed expression for the Pfaffian
(\ref{det4}).
The first values are
\begin{eqnarray*}
D_2^{(2)}(c) =& a(1+6a) \\ 
D_3^{(2)}(c) =& 8a^3(1+24a+45a^2+90a^3)\\ 
D_4^{(2)}(c) =&
1728a^6(1+60a+360a^2+2080a^3+2415a^4+2100a^5+2100a^6) \\
\end{eqnarray*}

It appears that $D_n^{(2)}(c)$ is always divisible by
$D_n^{(1)}(c)$ (which is not true for general Hankel
hyperdeterminants).
For $k>2$, we can compute the first polynomials:
\begin{eqnarray*}
D_2^{(3)}(c)=&a(1+30a+60a^2)\\
D_3^{(3)}(c)=&32a^3(1+240a+3285a^2+16650a^3+61425a^4\\&
+56700a^5+37800a^6)\\ D_2^{(4)}(c)=&a(1+126a+840a^2+840a^3)\\
D_3^{(4)}(c)=&
128a^3(1+2184a+134505a^2+1952370a^3+22027950a^4\\&+99542520a^5
+189552825a^6+246673350a^7\\&+ 130977000a^8+43659000a^9)
\end{eqnarray*}

It is also interesting to observe that the shifted determinants
$D_{n;r}^{(1)}(c)$ can be expressed as Wronskians
\begin{equation}
D_{n;r}^{(1)}(c)=a^{n(n-1)/2}W(b_r,b_{r+1},\ldots,b_{r+n-1})(a)\,.
\end{equation}
This identity follows immediately from the recursion
\begin{equation}
b_{n+1}(a)=a[b_n(a)+b'_n(a)]
\end{equation}
and is not of the same nature as the Karlin-Szeg\"o-type
identities like (\ref{KZbell}) below, discussed in section
\ref{orthopol}, in which the shifted Hankel determinant of order
$n$ is expressed in terms of a Wronskian of order $r$. Here, also,
$D_{n;r}^{(1)}(c)$ is always divisible by $D_n^{(1)}(c)$, as can
be checked from Table 1. In this case, the explanation is simple:
it follows from (\ref{detshift}) that
\begin{equation}
D_{n;r}^{(1)}(c)=\det(\<x^rP_i,P_j^*\>) D_n^{(1)}(c)
\end{equation}
where $P_j^*$ is the adjoint basis of $P_i$. The ratio
$D_{n;r}^{(1)}(c)/D_n^{(1)}(c)$ is therefore the determinant of
the operator $X^{(r)}_n=M_r\circ\pi_n$ where $M_r$ is
multiplication by $x^r$ and $\pi_n$ the orthogonal projection on
the subspace spanned by $P_0,\ldots,P_{n-1}$. The matrix elements
of $X^{(r)}_n$ can be read directly on the three-term recurrence
relation of the monic polynomials, by iterating it, if necessary,
to express $x^rP_i$ as a linear combination of the $P_j$'s. The
matrix element $\<x^rP_i,P_j^*\>$ is then equal to the coefficient
of $P_j$ in this expression.

For example, if the $P_n=C_n^{(a)}(x)$ are the monic Charlier
polynomials, the three-term recurrence is
\begin{equation}
xP_i=P_{i+1}+(i+a)P_i+iaP_{i-1}
\end{equation}
so that
\begin{equation}
\fl
x^2P_i=P_{i+2}+(2i+1+2a)P_{i+1}+(a^2+a+4ai+i^2)P_i+ia(2i-1+2a)P_{i-1}+i(i-1)aP_{i-2}
\end{equation}
and for $r=2$ and $n=3$, the matrix  is
\begin{equation}
X^{(2)}_3=\left[\matrix{ a+a^2 & a+2a^2 & 2a^2 \cr 1+2a & 1+5a+a^2
& 6a+4a^2\cr 1 & 3+2a & 4+9a+a^2 }\right]
\end{equation}
whose determinant is $6a^3+6a^4+3a^5+a^6$. This is the value at
$x=0$ of the $2\time 2$-Wronskian $W(C_3^{(a)},C_4^{(a)})(0)$ of
Charlier polynomials. There is a general formula (apparently new)
\begin{equation}\label{KZbell}
D_{n;r}^{(1)}(c)= (-1)^{rn} {W(C^{(a)}_n,\ldots
C_{n+r-1}^{(a)})(0) \over 1!2!\cdots (r-1)!} D_n^{(1)}(c)
\end{equation}
which will be derived in Section \ref{orthopol}.

\begin{table}
\caption{The first values of $D_{n;r}^{(1)}(c)$}
\begin{indented}
\item[]\begin{tabular}{@{}l|lll}
\br $n\backslash r$ & 0 & 1 & 2\\ \mr 1 & 1 & $a$ & $a+a^2$ \\ 2 &
$a$ & $a^3$ & $a^3(2+2a+a^2)$\\ 3 & $2a^3$ & $2a^6$ & $
2a^6(6+6a+3a^2+a^3)$\\ 4 & $12a^6$ &$12a^{10}$ &
$12a^{10}(24+24a+12a^2+4a^3+a^4)$\\ \br
\end{tabular}
\end{indented}
\end{table}

The sequence $D_2^{(k)}(c)$ of bidimensional hyperdeterminants
gives rise to an interesting triangle of integers $T_{k j}$
defined by
\begin{equation}
D_2^{(k)}(c)=\sum_{j=1}^k T_{k j}a^j
\end{equation}
whose first values are given in  Table 2.
\begin{table}
\caption{The triangle $T_{k j}$}
\begin{indented}
\item[]\begin{tabular}{@{}l|llllll}
\br $k\backslash j$ & 1 & 2 & 3 & 4 & 5 & 6 \\ \mr 1 & 1 &&&&&\\ 2
& 1 & 6 &&&&\\ 3 & 1 & 30 & 60 &&&\\ 4 & 1 & 126 & 840 & 840 &&\\
5 & 1 & 510 & 8820 & 25200 & 150120 &\\ 6 & 1 & 2046 & 84480 &
526680 & 831600 & 332640\\ \br
\end{tabular}
\end{indented}
\end{table}
The main diagonal is given by $(2k+1)!/k!$, and the second column
is $2^{2k-1}-2$. It is not difficult to give a generating function
for these numbers. According to (\ref{apolar}), if we know the
exponential generating function
\begin{equation}
g(x)=\sum_{n\ge 0} c_n {x^n\over n!}
\end{equation}
then
\begin{equation}\label{fg2a}
{1\over 2}+\sum_{k\ge 1} D_2^{(k)}(c) {x^{2k}\over (2k)!}
=
{1\over 2}g(x)g(-x)\,.
\end{equation}
Here, we have
\begin{equation}\label{fg2}
{1\over 2}+\sum_{k\ge 1} D_2^{(k)}(c) {x^{2k}\over (2k)!} =
{1\over 2}\exp[a(e^x+e^{-x}-2)]\,.
\end{equation}

Since the $b_n(a)$ are the moments of a discrete measure,
according to the general pattern, the calculation of
$D_{n;r}^{(k)}(c)$ amounts to sum the multiple series
\begin{equation}\label{serie}
D_{n;r}^{(k)}(c)= {e^{-na}\over n!}\sum_{m_1,\ldots,m_n\ge 0}
(m_1\cdots m_n)^r\Delta^{2k}(m_1,\ldots,m_n) {a^{m_1+\cdots
+m_n}\over m_1!\cdots m_n!}\,.
\end{equation}
But we can also express it as an integral for a continuous
measure. Indeed,
\begin{equation}
b_n(t)=e^{-t}\left(t{d\over dt}\right)^n e^t
\end{equation}
so that $(-1)^n e^{-t}b_n(-t)$ is the inverse Mellin transform of
$s^n\Gamma(s)$. We can then write
\begin{equation}\label{invmel}
(-1)^n e^{-a}b_n(-a)= {1\over 2\pi i}\int_{c-i\infty}^{c+i\infty}
s^n a^{-s}\Gamma(s)ds \quad (c>0)\,.
\end{equation}

We can choose $c=1$, and setting $s=1+iv$, we obtain
\begin{equation}
(-1)^n e^{-a}b_n(-a)= {1\over 2\pi a}\int_{-\infty}^\infty
(1+iv)^na^{-iv}\Gamma(1+iv)dv
\end{equation}
whence the integral representation of the hyperdeterminants
associated to $c'_n(a)=(-1)^n e^{-a}b_n(-a)$,
\begin{equation}\label{bellint}
D_n^{(k)}(c'(a))= {(-1)^{n(n-1)/2}\over (2\pi a)^n}\int_{\R^n}
\Delta^{2k}(v)\prod_{j=1}^na^{-iv_j}\Gamma(1+iv_j)dv_j\,.
\end{equation}
Since $D_n^{(k)}(c)$ is homogeneous of degree $n$ in the $c_i$ and
isobaric of weight $kn(n-1)$ (w.r.t. the weight function
$w(c_n)=n$), we have
\begin{equation}
D_n^{(k)}(c'(a))=e^{-na} D_n^{(k)}(c(-a))
\end{equation}
and comparing both expressions, we obtain the identity
\begin{equation}
\fl \int_{\R^n}
\Delta^{2k}(v)\prod_{j=1}^na^{-iv_j}\Gamma(1+iv_j)dv_j
=
(-1)^{n(n+1)\over 2}(2\pi a)^n\sum_{m_1,\ldots,m_n\ge 0}
\Delta^{2k}(m){(-a)^{m_1+\cdots+m_n}\over m_1!\cdots m_n!}
\end{equation}
which amounts to the evaluation of the  integral (\ref{bellint})
by the residue theorem. Since
\begin{equation}
a^{-iv}\Gamma(1+iv) =\exp\left\{ -i(\ln a +\gamma)v+\sum_{m\ge
2}\zeta(m){(-iv)^m\over m}\right\}
\end{equation}
it is tempting to make the choice $a=e^{-\gamma}$, in order to
cancel the linear term in the exponential, and to get a curious
identity involving on the left the values of the Riemann z\^eta
function at the integers, and on the right Euler's constant:
\begin{eqnarray}
\fl \int_{\R^n}\Delta^{2k}(x) \exp\left\{ \sum_{m\ge 2}
\zeta(m){p_m(-ix)\over m}\right\}dx_1\cdots dx_n \nonumber\\ \lo =
(-1)^{n(n-1)\over 2}(2\pi e^{-\gamma})^n \sum_{m_1,\ldots,m_n\ge
0}\Delta^{2k}(m) {(-e^{-\gamma})^{m_1+\cdots+m_n}\over m_1!\cdots
m_n!}\,.
\end{eqnarray}
Here, $p_m(x)=\sum_i x_i^n$ are the power-sums.

Let us now indicate an application of the above representation.
Instead of considering the shifted hyperdeterminants
$D_{n;r}^{(k)}$, which are the moments of the measure
$d\mu'(x)=x^rd\mu(x)$, one can replace $x^r$ by an arbitrary monic
polynomial $Q(x)$ of degree $r$. For a good choice of $Q(x)$, the
Hankel hyperdeterminants of the moments $c'_n$ of
$d\mu'(x)=Q(x)d\mu(x)$ may bear a simple relation to the original
ones.

In the case at hand, numerical experiments quickly suggest that
such a simple relation occurs only with the choice
\begin{equation}
Q(x)=(x)_r =x(x-1)\cdots (x-r+1)
\end{equation}
and that we have then
\begin{equation}\label{bQ}
D_n^{(k)}(c')=a^{nr}D_n^{(k)}(c)\,.
\end{equation}
This choice amounts to replace our original sequence $c_n=b_n(a)$
by
\begin{equation}
c'_n=b_n^{[r]}(a)=\sum_{j=0}^r s(r,j)b_{n+j}(a)
\end{equation}
where the $s(r,j)$ are the Stirling numbers of the first kind.
From (\ref{invmel}), we have
\begin{eqnarray}
c''_n&=e^{-a}b_n^{[r]}(-a) = {1\over 2\pi
i}\int_{c-i\infty}^{c+i\infty}
a^{-s}\Gamma(s)\sum_{j=0}^rs(r,j)(-s)^{n+j} ds\nonumber \\
&={1\over 2\pi i}\int_{c-i\infty}^{c+i\infty}
(-s)^na^{-s}\Gamma(s)(-s)(-s-1)\cdots(-s-r+1)ds\nonumber\\
&={(-1)^r\over 2\pi i}\int_{c-i\infty}^{c+i\infty}
(-s)^na^{-s}\Gamma(r+s)ds\,.
\end{eqnarray}
Hence,
\begin{eqnarray}
D_n^{(k)}(c'')& ={(-1)^r\over 2\pi
i}\int\cdots\int_{c-i\infty}^{c+i\infty}
\Delta^{2k}(s)\prod_{j=1}^n a^{-s_j}\Gamma(r+s_j)ds_j \nonumber\\
&={(-1)^r\over 2\pi i}\int\cdots\int_{c+r-i\infty}^{c+r+i\infty}
\Delta^{2k}(z)a^{rn}\prod_{j=1}^n
a^{-z_j}\Gamma(z_j)dz_j\nonumber\\
&=(-a)^{rn}D_n^{(k)}(e^{-a}b(-a))
\end{eqnarray}
which is equivalent to (\ref{bQ}).

\section{Miscellaneous examples}

\subsection{Hilbert hyperdeterminants}

Another classical example of a Hankel determinant which can be
evaluated in closed form is the Hilbert determinant
\begin{equation}
\left|{1\over i+j-1}\right|_{i,j=1}^n
=
D_n^{(1)}(c)
\end{equation}
where
\begin{equation}
c_n=\int_0^1 x^n dx\,.
\end{equation}
Thus, (\ref{kHeine}) gives immediately
\begin{equation}
\H(k,n):=D_n^{(k)}(c) ={1\over n!}S_n(1,1,k)
\end{equation}

In the simplest case $n=2$, the generating series (\ref{fg2})
gives
\begin{equation}
\H(k,2)={1\over(2k+1)(2k+2)}.
\end{equation}

Setting $c_n(a)=(n+a+1)^{-1}$, we obtain in the same way
\begin{equation}
D_n^{(k)}(c(a)) =\frac1{n!}S_n(1+a,1;k)
\end{equation}
For $a=r$, we obtain the hyperdeterminants $D_{n;r}^{(k)}(c)$.

\subsection{A class of Hankel-Wronskians}

In \cite{VD}, 4.12.3, one finds the Hankel determinants
$D_n^{(1)}(c)$ associated to the sequence
\begin{equation}
c_n={d^n\over dt^n} f(t) \quad {\rm with}\ f(t)=\left({e^t\over
1-e^t}\right)^x 
\end{equation}
whose particular case $x=1$ gives back one of the determinants
computed by Lawden \cite{Law}. To investigate the
hyperdeterminantal analogues, we will find it convenient to make
the substitution $t\rightarrow i\pi-t$, and to assume at first
that $-x=N$ is a positive integer. Up to a trivial sign, we can
now take $f(t)=(1+e^t)^N$, and our sequence is
\begin{eqnarray}
c_n &  ={d^n\over dt^n} (1+e^t)^N \nonumber \\ & = \sum_{k=0}^N
k^n {N\choose k} e^{kt} \nonumber \\ &= (1+e^t)^N\sum_{k=0}^N k^n
{N\choose k}\left({e^t\over 1+e^t}\right)^k
                                \left(1-{e^t\over 1+e^t}\right)^{N-k}\nonumber\\
&= (1-p)^{-N}\sum_{k=0}^N k^n {N\choose k}p^k(1-p)^{N-k}
\label{binodis}
\end{eqnarray}
where $p=e^t/(1+e^t)$. That is, $(c_n)$ is the moment sequence of
the binomial distribution, for which the Krawtchouk polynomials
$K_n(x;p,N)$ are orthogonal. We have
\begin{equation}
\sum_{k=0}^N {N\choose k}p^k(1-p)^{N-k} K_m(k) K_n(k) = {(-1)^n
n!\over (-N)_n}\left({1-p\over p}\right)^n \delta_{mn}
\end{equation}
whilst the monic polynomials $\tilde K_n$ are related to the
standard ones by
\begin{equation}
\tilde K_n(x) = p^n (-N)_n K_n(x) \,.
\end{equation}
Hence,
\begin{equation}
\| \tilde K_n\|^2=(-1)^n n! p^n(1-p)^n (-N)_n
\end{equation}
which gives for the Hankel determinant
\begin{eqnarray}
D_n^{(1)}(c)&=(1-p)^{-Nn}\prod_{j=0}^{n-1}\| \tilde
K_j\|^2\nonumber\\ &={(-e^t)^{n(n-1)/2}\over (1+e^t)^{n(-N+n-1)}}
\prod_{j=0}^{n-1}j! (-N)_j
\end{eqnarray}
which agrees with  Theorem 4.59 of \cite{VD} after substituting
back $t\rightarrow i\pi-t$ and $N=-x$, and can be extended as
usual to values of $x$ ranging over the whole complex plane by
means of Carlson's theorem.

The intepretation in terms of the Krawtchouk polynomials allows
one to go one step further and to find a closed form for the
$D_{n;1}^{(1)}$. Indeed, we know that
\begin{equation}
D_{n;1}^{(1)}= \det(X_n) D_{n}^{(1)}
\end{equation}
where $X_n$ is the operator of multiplication by $x$ followed by
the orthogonal projection on the subspace spanned by the first $n$
Krawtchouk polynomials. The matrix of $X_n$ can be read directly
on the three-term recurrence
\begin{equation}
\fl x\tilde K_n(x)= \tilde K_{n+1}(x) +[p(N-n)+n(1-p)]\tilde
K_n(x) +np(1-p)(N+1-n)\tilde K_{n-1}(x)
\end{equation}
which yields the tridiagonal matrix
\begin{equation}\fl
X_n=\left[\matrix{ \beta          &            1    & 0 & 0 &
\cdots & 0 & 0 \cr \lambda(\mu-1) & \alpha+\beta    & 1 & 0 &
\cdots & 0 & 0 \cr 0              &2\lambda(\mu-2)  &
2\alpha+\beta & 1 & \cdots & 0 & 0 \cr \vdots         & &  \ddots
&   &        &   &\vdots\cr 0              & 0 & 0             & &
\cdots &(n-1)\lambda(\mu-n+1)&(n-1)\alpha+\beta} \right]
\end{equation}
where $\alpha=1-2p$, $\beta =Np$, $\lambda=p(1-p)$ and $\mu=N+1$.
The three-term recurrence for the tridiagonal determinants is
easily solved by means of a generating function, and one finds
\begin{equation}
\det(X_n) =(N)_np^n\,.
\end{equation}

The other shifted determinants $D_{n;r}^{(1)}$ can in principle be
calculated by the same method, but is does not seem possible to
solve the recurrences in closed form, and indeed, numerical
calculations show that no nice factorised expression can be
expected, except in the special case $r=2$ and $N=-1$, which gives
back another one of Lawden's determinants \cite{Law}. For the
operator $X^{(2)}_n$, multiplication by $x^2$ followed by
projection, we obtain, for $N=-1$,
\begin{equation}
\det(X^{(2)}_n)=(n!)^2e^{nt}{e^{(n+1)t}-(-1)^{n+1}\over
(1+e^t)^{2n+1}}
\end{equation}
but no other case seems to lead to an interesting formula. But
this does not rule out the possibility of replacing $x^r$ by
another monic polynomial $Q(x)$ of degree $r$.

To investigate this possibility, we shall  adopt the same strategy
as in the case of Bell polynomials, and  look for an integral
representation of our sequence. Once again, it will be convenient
to work with a slightly modified (but equivalent) sequence
\begin{equation}
a_n={d^n\over dt^n} g(t) \quad {\rm where}\
g(t)=(1-e^{-t})^N=f(i\pi-t)\,.
\end{equation}
The Laplace transform of $g(t)$ is
\begin{eqnarray}
G(s) &= \int_0^\infty e^{-st}(1-e^{-t})^Ndt =
\int_0^1u^{s-1}(1-u)^Ndu\nonumber\\ &= \Beta(s,N+1)= {N!\over
s(s+1)\cdots(s+N)}\,.
\end{eqnarray}
Hence,
\begin{equation}
a_n = {1\over 2\pi i}\int_{c-i\infty}^{c+i\infty} s^n
{N!e^{ts}\over s(s+1)\cdots(s+N)} ds
\end{equation}
and
\begin{equation}\label{eq61}
D_{n;r}^{(k)}(a) = {1\over n!(2\pi
i)^n}\int\cdots\int_{c-i\infty}^{c+i\infty}
\Delta^{2k}(s)\prod_{j=1}^n{N! s_j^r e^{ts_j} ds_j\over
s_j(s_j+1)\cdots (s_j+N)}\,.
\end{equation}
Remark that by (\ref{binodis}), we know that $D_{n;r}^{(k)}(a)$ is
also equal to the finite sum
\begin{equation}\fl
D_{n;r}^{(k)}(a)={1\over n!}\sum_{k_1,\ldots,k_n=0}^N
\Delta^{2k}(k_1,\ldots,k_n)(k_1\cdots
k_n)^r(-e^{-t})^{k_1+\cdots+k_n} {N\choose k_1}\cdots {N\choose
k_n}
\end{equation}
which is indeed the value of the integral (\ref{eq61}) according to the residue
theorem.

From (\ref{eq61}), we see that if we replace $(a_n)$ by the new
sequence
\begin{equation}
a'_n=
 {1\over 2\pi i}\int_{c-i\infty}^{c+i\infty} s^n Q(s) {N!e^{ts}\over s(s+1)\cdots(s+N)} ds
\end{equation}
where $Q(s)={(s)_r\over r!}$, the hyperdeterminants
\begin{equation}
d^{(k)}_n(r;N)=D^{(k)}_n(a')
\end{equation}
satisfy to
\begin{equation}
d^{(k)}_n(r;N)={N\choose r}^n e^{-nrt}d^{(k)}_n(0;N-r)
\end{equation}
that is, are expressible in terms of the unshifted
hyperdeterminants of the original sequence, with parameter $N-r$.

As in the case of Bell polynomials, the $D_n^{(2)}(c)$ can be
reduced to a Pfaffian.

%
%
%
%

\section{Examples involving orthogonal polynomials}\label{orthopol}

Hankel determinants associated to sequences of the form
$c_n=Q_n(x)$, where $(Q_n)$ is a family of orthogonal polynomials,
have been called {\it Tur\'anians} by Karlin and Szeg\"o, who
computed their values for the classical families \cite{KS}. Recent
references on this subject can be found in \cite{Le}, where these
results have been generalized by a different method based on a
little-known determinantal identity due to Turnbull.

In this section, we will calculate the hyperdeterminantal
analogues of the Tur\'anians evaluated in \cite{KS}. As in the
preceding section, we will make use of  the integral
representations of the classical orthogonal polynomials.
Interestingly enough, Selberg's formula will not be sufficient to
deal with these cases, and we will have to rely upon one of its
extensions, which is due to Kaneko \cite{Ka}.

\subsection{Kaneko's integral and its variants}

The required integral formula involves the {\it generalized Jacobi
polynomials} $p_\kappa^{\alpha,\beta,\gamma}(y)$
\cite{Ko,V,D,La1}, which are the symmetric polynomials in $r$
variables $(y_1,\ldots,y_r)$ obtained by applying the Gram-Schmidt
process to the basis of monomial symmetric functions $m_\mu(y)$
(ordered by the condition $\mu < \nu$ if $|\mu|<|\nu|$, or
$|\mu|=|\nu|$ and $\mu$ precedes $\nu$ for the reverse
lexicographic order) with respect to the measure
\begin{equation}
d\mu^{\alpha,\beta\gamma}(y) =|\Delta(y)|^{2\gamma+1}\prod_{i=1}^r
(1-y_i)^\alpha(1+y_i)^\beta dy_1\cdots dy_r
\end{equation}
on $[-1,1]^r$, normalized by the condition that the leading term
of $p_\kappa^{\alpha,\beta,\gamma}(y)$ is $m_\kappa(y)$. 
Let 
\begin{equation}
R(x,y)=\prod_{i=1}^n\prod_{j=1}^r(x_i-y_j) \,.
\end{equation}
Kaneko's formula reads
\begin{eqnarray}\label{Kaneko}
\fl \int_{[0,1]^n} 
R(x,y)
\prod_{i=1}^n x_i^{a-1} (1-x_i)^{b-1} |\Delta(x)|^{2c} dx_1\cdots
dx_n \nonumber\\ 
\lo\qquad =
2^{-nr}S_n(a,b;c)p_{(n^r)}^{\alpha,\beta,\gamma}(1-2y_1,\ldots,1-2y_r)
\end{eqnarray}
where $\alpha={a\over c}-1$, $\beta={b\over c}-1$ and
$\gamma={c}-{1\over 2}$.

The multivariate Jacobi polynomials indexed by rectangular
partitions can be expressed in terms of generalized hypergeometric
functions. This expression is simpler in terms of the polynomials
\begin{equation}\label{JacobiP}
P_\kappa^{(a,b)}(y_1,\ldots,y_r;\case1c)
=
{p_\kappa^{a,b,c-{1 \over 2}}(1-2y_1,\ldots,1-2y_r) \over
p_\kappa^{a,b,c-{1 \over 2}}(1,\ldots,1)}
\end{equation}
which are orthogonal on $[0,1]^r$ for the Selberg measure with
parameters $(a+1,b+1,c)$. For a rectangular partition
$\kappa=(n^r)$,
\begin{equation}\label{Pas2F1}
P_{(n^r)}^{(a,b)}(y_1,\ldots,y_r;\case1c)
=
{}_2F_1^{(1/c)}\left( \left.\matrix{-n\, ;\, a+b+s+n\cr
a+s}\right| y_1,\ldots,y_r\right)
\end{equation}
where  $s=1+(r-1)c$. The generalized
hypergeometric functions associated with Jack polynomials
$C_\kappa^{(\alpha)}(y_1,\ldots,y_r)$
 are defined by
\cite{Ka,Ko}
\begin{equation}
\fl {}_pF_q^{(\alpha)}\left( \left.\matrix{a_1\ldots a_p\cr
b_1\ldots b_q}\right| y_1,\ldots,y_r\right)
=
\sum_{n\ge0}{1 \over n!}\sum_{|\kappa|=n} {
[a_1]^{(\alpha)}_\kappa \cdots [a_p]^{(\alpha)}_\kappa \over
[b_1]^{(\alpha)}_\kappa \cdots [b_q]^{(\alpha)}_\kappa
}C_\kappa^{(\alpha)}(y_1,\ldots,y_r)
\end{equation}
where
\begin{equation}
[a]^{(\alpha)}_\kappa =\prod_{i=1}^{\ell(\kappa)}\left(a-{1
\over\alpha}(i-1)\right)_{\kappa_i}\,.
\end{equation}
We note that in the case $\kappa=(n^r)$, the denominator of
(\ref{JacobiP}) is given by (\ref{Kaneko}) as
\begin{equation}
p_\kappa^{a,b,c-{1 \over 2}}(1,\ldots,1) = 2^{nr} {S_n(a,b+r,c)\over
S_n(a,b,c)}\,.
\end{equation}
This formula is needed to calculate the degenerate cases of
Kaneko's integral corresponding to Laguerre and Hermite
polynomials.

The generalized Laguerre polynomials $L_\kappa^a(y;\alpha)$ are
defined by \cite{La2} (see also \cite{BF})
\begin{equation}\label{deflag}
L_\kappa^a(y_1,\ldots,y_r;\alpha)
=
\lim_{b\rightarrow\infty} P_\kappa^{(a,b)} \left( {y_1\over
b},\ldots,{y_r\over b};\alpha\right) 
\end{equation}
(we use there the convention of \cite{Kli}).
Let $LS_n(a,c)$ denote the Laguerre-Selberg integral (\ref{LagSel}). 
One can deduce from (\ref{Kaneko})  the Laguerre version of
Kaneko's integral. Indeed, Kaneko's formula can also be written as
\cite{Ka}
\begin{eqnarray}\fl
\int_{[0,1]^n}R(x,y)|\Delta(x)|^{2c}\prod_{i=1}^nx_i^{a-1}(1-x_i)^{b-1}dx_i
\nonumber\\
=S_n(a+r,b,c)
{}_2F_1^{(c)}\left(\left.-n;\frac1c(a+b+r-1)+n-1\atop\frac1c(a+r)-1\right|y_1,\cdots,y_r\right)
\end{eqnarray}
Setting $x_i=u_i/L$, $y_i=v_i/L$ and letting $L\rightarrow\infty$
in this formula we obtain
\begin{eqnarray}\fl
\int_{(a,\infty)^n}R(u,v)|\Delta(u)|^{2c}\prod_{i=1}^nu_i^{a-1}e^{-u_i}du_i\nonumber\\
= \lim_{L\rightarrow\infty}L^{nr+cn(n-1)+n(a-1)+n}S_n(a+r,L+1,c)\times\nonumber\\
\quad \times{}_2F_1^{(c)}
\left(\left.-n,\frac1c(a+r+L)+n-1\atop
\frac1c(a+r-1)\right|{v_1\over L},\cdots,{v_l\over L}\right)
\end{eqnarray}
From (\ref{Pas2F1}) and (\ref{deflag}) we have, setting  $b'={L+1\over c}-1$,
\begin{eqnarray}\fl
\lim_{L\rightarrow\infty}{}_2F_1^{(c)}\left(\left.-1,\frac1c(a+r+L)+n-1\atop
\frac1c(a+r-1)\right|{v_1\over L},\cdots,{v_r\over
L}\right)\nonumber\\
= \lim_{b'\rightarrow
\infty}{}_2F_1^{(c)}\left(\left.-1,a'+b'+s'+n\atop
a'+s'\right|{v_1\over cb'+c-1},\cdots,{v_r\over
cb'+c-1}\right)\nonumber\\ =L^{a'}_{n^r}({v_1\over
c},\cdots,{v_r\over c};c)
\end{eqnarray}
where $a'=\frac ac-1$ and $s'=1+{r-1\over c}$.

On another hand,
\begin{equation}
\lim_{L\rightarrow\infty}L^{nr+cn(n-1)+n(a-1)+n}S_n(a+r,L+1;c)=LS_n(a+r,c)
\end{equation}
Finally, we get
\begin{equation}\label{LagKan}
\fl \int_{(0,\infty)^n}
R(x,y)
\Delta^{2k}(x)\prod_{i=1}^nx_i^{a-1}e^{-x_i}dx_i =LS_n(a+r,c)
L_{(n^r)}^{{a\over c}-1}\left( {y_1\over c},\ldots,{y_r\over
c};c\right)\,.
\end{equation}

 Similarly,
an appropriate limit of
(\ref{Kaneko}) yields
\begin{eqnarray}\label{KanekoHermite}
\fl \int_{\R^n}
R(x,y)
\Delta^{2k}(x)\prod_{i=1}^n e^{-x_i^2}dx_i =
(-1)^{nr\over2}\pi^{n\over2}2^{-\frac12kn(n-1)-nr}k^{nr\over2}&\prod_{j=1}^n{(kj)!\over
k!}\times\nonumber\\ 
\lo\qquad\times H_{(n^r)}\left(i{y_1\over\sqrt
k},\cdots,i{y_r\over\sqrt k};k\right)
\end{eqnarray}
where the generalized Hermite polynomials $H_\kappa(y;\alpha)$ are
defined by
\begin{equation}\fl
H_\kappa(y_1,\cdots,y_r;\alpha)= \lim_{a\rightarrow\infty}
(-\sqrt{2a})^{|\kappa|}L_\kappa^a(a+y_1\sqrt{2a},\cdots,a+y_r\sqrt{2a};\alpha)\,.
\end{equation}
We follow here the convention of \cite{BF}.

Kaneko's identity can be
interpreted as a generalization of Heine's integral representation
of orthogonal polynomials in the Jacobi case. Indeed, it can be
rewritten as
\begin{equation}
p_{(n^r)}^{\alpha,\beta,\gamma}(t_1,\ldots,t_r) ={1\over
Z_n^{\alpha,\beta,\gamma}} \int_{[-1,1]^n}
|\Delta(x)|^{2c}d\mu_t(x_1)\cdots d\mu_t(x_n)
\end{equation}
where $d\mu_t(x)=\prod_{j=1}^r(t_j-x)(1-x)^{a-1}(1+x)^{b-1}$,
$Z_n^{\alpha,\beta,\gamma} = 2^{cn(n-1)+n(a+b+r-2)}S_n(a,b,c)$,
$a=c(\alpha+1)$, $b=c(\beta+1)$, $c=\gamma+\case12$.

Hence, when $c=k$ is a positive integer, the  symmetric Jacobi
polynomials indexed by rectangular partitions are expressible as
hyperdeterminants
\begin{equation}
Z_n^{\alpha,\beta,\gamma}
p_{(n^r)}^{\alpha,\beta,\gamma}(t_1,\ldots,t_r) =n! D_n^{(k)}(c(t))
\end{equation}
where
\begin{equation}
c_m(t)=\int_{-1}^1 x^m d\mu_t(x)\,.
\end{equation}
This can be regarded as a generalization of the classical
determinantal expression of the orthogonal polynomials in terms of
the moments.

The extension of these identities to non-rectangular partitions
or to other measures appears to be unknown.

\subsection{The case $k=1$: Leclerc's identity}

On another hand, for $k=1$, Kaneko's representation can be
extended to general orthogonal polynomials. Let $\mu$ be any
linear functional such that the bilinear form $(f,g)=\mu(fg)$ is
non degenerate, and extend it as above to functions of $n$
variables $x_1,\ldots,x_n$ by setting $\mu_n(f_1(x_1)\cdots
f_n(x_n))=\mu(f_1)\cdots\mu(f_n)$. Let
$p_\lambda^{(k)}(x_1,\ldots,x_n)$ be the basis of symmetric
polynomials obtained by applying the Gram-Schmidt process to the
monomial basis with respect to the scalar product 
\begin{equation}
\<f,g\>_k=\mu_n\left( \Delta^{2k}(x) f(x)g(x)\right)
\end{equation}
with leading term $m_\lambda$. In this section, we shall use the
representation of partitions by {\it weakly increasing} sequences
$\lambda=(0\le \lambda_1\le\ldots \le\lambda_n)$ instead of the
usual one (this will be more convenient for the indexing of
minors). The one-variable polynomials $p_m(x)$ are  the monic
orthogonal polynomials associated with $\mu$.

When $k=1$, one has
\begin{equation}\label{orth1}
p_\lambda^{(1)}(x)={D_\lambda(x)\over\Delta(x)}
\end{equation}
where the alternants (Slater determinants)
$D_\lambda(x)=\det(p_{\lambda_i+i-1}(x_j))$ form the natural basis
of antisymmetric orthogonal polynomials for $\mu_n$. Now, 
we can write
\begin{equation}
R(y,x)=\prod_{j=1}^r\prod_{i=1}^n(y_j-x_i)
={\Delta(x,y)\over\Delta(x)\Delta(y)}\,.
\end{equation}
The analogue of Kaneko's integral in this context is the
scalar product
\begin{equation}
\mu_n(\Delta^2(x)R(y,x))=\<1,R(y,x)\>_1\,.
\end{equation}
Expressing $\Delta(x,y)$ in terms of the one-variable monic
orthogonal polynomials $p_m$ as
\begin{equation}\label{dbvd}
\Delta(x,y)=\det(p_{i-1}(x_j)|p_{i-1}(y_k))
\end{equation}
and taking the Laplace expansion of  this determinant of order
$n+r$ with respect to its first $n$ columns (containing the
variables $x_i$), we find that
\begin{equation}\fl
R(y,x)=
{1\over\Delta(x)\Delta(y)}\sum_{\alpha,\beta}(-1)^{|\alpha|}D_\alpha(x)D_\beta(y)
=\sum_{\alpha,\beta}(-1)^{|\alpha|}p_\alpha(x)p_\beta(y)
\end{equation}
where the sum runs over all pairs of partitions
\begin{equation}
\alpha=(0\le \alpha_1\le \ldots\le\alpha_n)\,,\quad \beta=(0\le
\beta_1\le \ldots\le\beta_r)
\end{equation}
such that
$(\alpha_1+1,\ldots,\alpha_n+n,\beta_1+1,\ldots,\beta_r+r)$ is a
permutation of $(1,2,\ldots,n+r)$, in which case $(-1)^{|\alpha|}$
is its sign. In particular, for $\alpha=0$, $\beta$ is the
rectangular partition $\beta=(n^r)$, so that

\begin{eqnarray}
\<1,R(y,x)\>_1& = \sum_{\alpha,\beta} (-1)^{|\alpha|}
p^{(1)}_{\beta}(y)\<p^{(1)}_0 , p^{(1)}_\alpha\>\nonumber \\
&=\mu_n(\Delta^2(x)) p^{(1)}_{(n^r)}(y_1,\ldots,y_r)\,.
\end{eqnarray}
This equation contains as a special case Theorem 1 of \cite{Le},
which in turn implies all the identities of Karlin and Szeg\"o as
well as many other ones. Indeed, taking the limit $y_i\rightarrow
u$, $i=1,\ldots,r$ in (\ref{orth1}), we obtain a Wronskian of
one-variable polynomials
\begin{equation}
p_\lambda^{(1)}(u,\ldots,u)=
{W(p_{\lambda_1},p_{\lambda_2+1},\ldots,p_{\lambda_r+r-1})(u)
\over 1!2!\cdots (r-1)!}
\end{equation}
(cf. \cite{Meh2}, 7.1.1 p. 107), so that
\begin{equation}\label{BLA}
\mu_n\left(\Delta^2(x)\prod_{i=1}^n(y-x_i)^r\right) =
\mu_n(\Delta^2) {W(p_n,\ldots,p_{n+r-1})(y) \over 1!2!\cdots
(r-1)!} \,.
\end{equation}
But we have also
\begin{equation}\label{BLB}
\mu_n\left(\Delta^2(x)\prod_{i=1}^n(y-x_i)^r\right)= (-1)^{nr} n!
\det(c_{r+i+j}(y))|_0^{n-1}
\end{equation}
where
\begin{equation}
c_m(y)=\mu\left[ (x-y)^m\right]=  \sum_{j=0}^m{m\choose
j}\mu(x^j)(-y)^{m-j}\,.
\end{equation}
The equality of the right-hand sides of (\ref{BLA}) and
(\ref{BLB}) is precisely Theorem 1 of \cite{Le}.

Applying this identity (with $y=0$) to the case where the moments
$c_n$ are the Bell polynomials $b_n(a)=\mu(x^n)$,
so that $P_n(x)=C_n^{(a)}(x)$, we obtain (\ref{KZbell}).

This suggest the conjecture that in general
$\mu_n(\Delta^{2k}(x)R(y,x))$ should be expressible as
$\mu_n(\Delta^{2k}(x)) q_{(n^r)}^{(k')}(y)$, where the
$q_\lambda^{(k')}$ are the symmetric orthogonal polynomials for
another functional $\mu'$ related to $\mu$ in some natural way.

\subsection{Hypertur\'anians of Legendre polynomials}

Let us start, as in \cite{KS}, with the Legendre polynomials
$P_n(x)$. Laplace's integral representation
\begin{equation}
P_n(x)={1\over\pi}\int_0^\pi (x+\cos\phi\sqrt{x^2-1})^nd\phi
\end{equation}
can be rewritten as
\begin{equation}
P_n(x)=\int_a^b t^n d\mu(t)
\end{equation}
where $a=x-\sqrt{x^2-1}$, $b=x+\sqrt{x^2-1}$, and
$d\mu(t)=\pi^{-1}(t-a)^{-1/2}(b-t)^{-1/2}$. Hence,
\begin{equation}
D_n^{(k)}(P(x))
={1\over n!\pi^n}\int_{[a,b]^n}\Delta^{2k}(t)
\prod_{i=1}^n(t_i-a)^{-1/2}(b-t_i)^{-1/2}dt_i
\end{equation}
which under the substitution $t_i=(b-a)u_i+a$ reduces to the
Selberg integral
\begin{eqnarray}\fl
D_n^{(k)}(P(x))= {1\over n!\pi^n}(b-a)^{kn(n-1)} \int_{[0,1]^n}
\Delta^{2k}(u)\prod_{i=1}^nu_i^{-1/2}(1-u_i)^{-1/2}du_i\nonumber \\
\lo={1\over n!\pi^n}(2\sqrt{x^2-1})^{kn(n-1)} S_n(\case12,
\case12;k)\,.
\end{eqnarray}

Now, the shifted polynomials $c_n=P_{r+n}(x)$ are the moments
\begin{equation}
P_{r+n}(x)=\int_a^b t^n d\mu_r(t)
\end{equation}
where $d\mu_r(t)=\pi^{-1}t^r(t-a)^{-1/2}(b-t)^{-1/2}dt$, so that
\begin{eqnarray}
\fl D_{n;r}^{(k)}(P(x))
= {1\over
n!\pi^n}\int_{[a,b]^n}\Delta^{2k}(t) (t_1\ldots t_n)^r
\prod_{i=1}^n(t_i-a)^{-1/2}(b-t_i)^{-1/2}dt_i\nonumber\\
\lo=
 {(b-a)^{kn(n-1)+rn}\over n!\pi^n}
\int_{[0,1]^n} \Delta^{2k}(u) \prod_{i=1}^n(u_i+v)^r
u_i^{-1/2}(1-u_i)^{-1/2}du_i
\end{eqnarray}
where $v={a\over b-a}$. This is of the form (\ref{Kaneko}) with
$y_1=y_2=\ldots=y_r=-v$, whence, since
$1+2v={x\over\sqrt{x^2-1}}$,
\begin{eqnarray}
\fl D_{n;r}^{(k)}(P(x))=&
2^{kn(n-1)}(x^2-1)^{\frac12(kn(n-1)+nr)}
{1\over
n!\pi^n}S_n(\case12,\case12;k)\times\nonumber\\&\qquad \times
P_{(n^r)}^{\alpha,\beta,\gamma}\left(
{x\over\sqrt{x^2-1}},\ldots,{x\over\sqrt{x^2-1}}\right)
\end{eqnarray}
where $\alpha={1\over 2k}-1$. $\beta={1\over 2k}-1$ and $\gamma=k-\case12$.

It is instructive to have a look at the case $k=1$. Here,
$\alpha=\beta=-{1\over 2}$ and  the generalized Jacobi
polynomials are the symmetric orthogonal polynomials for the
measure
\begin{equation}\label{mes1}
d\mu(y)=\Delta^2(y) \prod_{i=1}^r {dy_i\over\sqrt{1-y_i^2}}\,.
\end{equation}
For $r=1$, the orthogonal polynomials are the Chebyshev
polynomials $T_n(y)$, and 
the symmetric orthogonal polynomials for (\ref{mes1}) are
the $D_\mu(y)/\Delta(y)$ formed from
corresponding monic polynomials $t_m$. Taking the limit of
these expressions for $(y_1,\ldots,y_r)\rightarrow
(\xi,\ldots,\xi)$, where $\xi={x\over\sqrt{x^2-1}}$, we obtain a
Wronskian of Chebyshev polynomials evaluated at $\xi$,
which is precisely the expression of the Tur\'anian found by Karlin and Szeg\"o 
(see also \cite{Le}).

\subsection{Laguerre }

We start with the hypergeometric representation of the monic
Laguerre polynomials
\begin{equation}
\tilde L_n^{(a)}(x)={}_1F_1\left(\left.-n\atop
a+1\right|x\right)=\lim_{b\rightarrow\infty}{}_2F_1\left(\left.-n,b\atop
a+1\right|{x\over b}\right)
\end{equation}
The second part of this equality leads to write each shifted
hypertur\`anian as the limit of a Kaneko integral, which gives
after simplification
\begin{eqnarray}
\fl D_{n;r}^{(k)}(\tilde L^{(a)})=& {1\over n!k!^n}
\lim_{b\rightarrow\infty}{\left(-x\over b\right)^{kn(n-1)+nr}}
\prod_{j=0}^{n-1}{(b)_{jk+r}(a-b+1)_{jk}(jk+k)!\over(a+1)_{k(n+j-1)+r}}
\times\nonumber\\ 
&\qquad\times P_{n^r}^{\frac bk-1,{a-b+1\over k}-1}(\frac
bx,\cdots,\frac bx;k)
\end{eqnarray}
From  (\ref{Pas2F1}), we see that this can be written as a
generalized hypergeometric function
\begin{eqnarray}
D_{n;r}^{(k)}(\tilde L^{(a)})&=
(-1)^{{kn(n-1)\over 2}+nr}
x^{kn(n-1)+nr}{1\over n!k!^n}
\prod_{j=0}^n{(jk+k)! \over
(a+1)_{k(n+j-1)+r}}\times\nonumber\\
&\qquad\times\ _2F_0^{(k)}\left(\left. -n,\
{a+r\over k}+n-1\atop -\right|\frac kx,\cdots,\frac
kx\right)\end{eqnarray} 
In particular, if $r=0$, we obtain
\begin{equation}
D_n^k(\tilde L^{(a)}) =(-1)^{kn(n-1)\over 2}x^{kn(n-1)} {1\over n!k!^n}
\prod_{j=0}^{n-1}{(jk+k)!\over (a+1)_{k(n+j-1)}}
\end{equation}

\subsection{Hermite}

We start as above with the representation of the monic Hermite polynomials
as  limits of  hypergeometric functions
 \begin{equation} 
\tilde H_n(x)=\lim_{a\rightarrow\infty}a^{n\over2}\
_2F_1\left(\left.-n,2a\atop a\right|\frac12\left(1-\frac x{\sqrt
a}\right)\right)\,.
\end{equation}
We can then write the shifted hypertur\'anian as the limit of a
Kaneko integral.

If $r>0$, from (\ref{KanekoHermite}), one finds

\begin{eqnarray}
\fl D_{n,r}^{(k)}(\tilde
H)={(-1)^{\frac12kn(n-1)}2^{-\frac12kn(n-1)-nr}k^{nr\over 2}}{1\over
n!k!^n}\prod_{j=1}^n(jk)!
H_{(n^r)}({x\over\sqrt k},\cdots,{x\over \sqrt k};k).
\end{eqnarray}

 In the simplest case  $r=0$, we obtain
\begin{equation}
D_{n}^{(k)}(\tilde H)={(-{1\over2})^{\frac12kn(n-1)}}{1\over
n!k!^n}\prod_{j=1}^n(jk)!\,.
\end{equation}

Let us remark that this calculation is connected to the evaluation
of $\left(\sum_{j=1}^n{\partial^2\over\partial
x_j^2}\right)^{N}\Delta(x)^{2k}$, 
($N={kn(n-1)\over 2})$, which can be found
in \cite{Meh1} (17.6.9). Indeed, expanding $\Delta(x)^{2k}$, one
has
\begin{eqnarray}
\fl \left(\sum_{j=1}^n{\partial^2\over\partial
x_j^2}\right)^{N} \Delta(x)^{2k}&=\sum_{\sigma_1,\cdots,\sigma_{2k}\in\SG_n}
\epsilon(\sigma_1)\cdots\epsilon(\sigma_{2k})\times\nonumber\\
&\quad\times\left(\sum_{l_1,\cdots,l_k} \left({N}\atop l_1\cdots
l_n\right) \prod_{j=1}^n{\partial^{2l_j}\over\partial x_j^{2l_j} }
\right)\prod_{i=1}^nx_i^{\sigma_1(i)+\cdots+\sigma_{2k}(i)-2k}
\end{eqnarray}
But for each monomial $x_1^{p_1}\cdots x_n^{p_n}$ appearing in the
previous formula, we obtain
\begin{eqnarray}
\left(\sum_{l_1,\cdots,l_k} \left({N}\atop l_1\cdots
l_n\right)\right.&\left.
\prod_{j=1}^n{\partial^{2l_j}\over\partial x_j^{2l_j} }
\right)\left.\prod_{i=1}^nx_i^{p_1}\right|_{x_i=0}\nonumber \\
&=\left\{
\begin{array}{ll}
N!\prod_{i=1}^n{p_i!\over{p_i\over2}!}&\mbox{if
each $p_i$ is even}\\ 0&\mbox{otherwise} .
\end{array}
\right.
\end{eqnarray}
It follows that
\begin{eqnarray}
\left(\sum_{j=1}^n{\partial^2\over\partial
x_j^2}\right)^{N}\Delta(x)^{2k}&=(-1)^{N}n!2^{2N}
N!
D_n^{(k)}(\tilde H)\\&={2^{N}N!\over
k!^n}\prod_{j=1}^n(jk)!
\end{eqnarray}

\subsection{Charlier }

The monic Charlier polynomials $C_n^{(a)}(x)=n!L_n^{(x-n)}(a)$ are
given by the exponential generating function
\begin{equation}
\sum_{n\ge 0} C_n^{(a)}(x) {t^n\over n!}
=
e^{-at}(1+t)^x\,.
\end{equation}

One has the integral representation (see \cite{Kar} p. 446)
\begin{equation}
C_n^{(a)}(x)={1\over\Gamma(-x)} \int_0^\infty
e^{-t}t^{-x-1}(t-a)^n dt\,.
\end{equation}

From this, we get easily the Hankel hyperdeterminants associated
to the sequence $c_n=C_n^{(a)}(x)$. The result can be cast in the
form
\begin{equation}
D_n^{(k)}(c)= {1\over
n!}\prod_{j=0}^{n-1}{(k+kj)!\, \Gamma(-x+kj)\over k!\, \Gamma(-x)}\,.
\end{equation}
In particular, for the determinants,
\begin{equation}
D_n^{(1)}(c)= \prod_{j=0}^{n-1} {\Gamma(j+1)\Gamma(j-x)\over
\Gamma(-x)}\,.
\end{equation}

The shifted hyperdeterminants $D_{n;r}^{(k)}(c)$ can be similarly
evaluated via Kaneko's identity in the Laguerre form, 
\begin{equation}\fl
D_{n;r}^{(k)}(c)={1\over
n!k!^n}\prod_{j=0}^{n-1}(-x)_{jk+r}(jk+k)!L_{n^r}^{-\frac xk-1}
\left(\frac ak,\cdots,\frac ak;k\right)\,.
\end{equation}
For the shifted determinants, we get a Wronskian of Laguerre
polynomials
which is easily seen to be equivalent to the evaluation given in
\cite{KS}.

\subsection{Meixner}

The Meixner polynomials admit the integral representation
(\cite{Kar} p. 448)
\begin{equation}
\fl \phi_n(-x;\beta,\gamma)
=
{\Gamma(\beta)\over \Gamma(\beta-x)\Gamma(x)} \int_0^1
t^{x-1}(1-t)^{\beta-x-1} \left[ 1+\left( {1\over
\gamma}-1\right)t\right]^n dt\,.
\end{equation}
From this, one deduces the hyperdeterminants associated to
$c_n=\phi(-x;\beta,\gamma)$
\begin{equation}
\fl D_n^{(k)}(c)= {1\over
n!}\left({1-\gamma\over\gamma}\right)^{nk(n-1)} \prod_{j=0}^{n-1}
{\Gamma(x+jk)\Gamma(\beta-x+jk)\Gamma(\beta)\Gamma(jk+k+1) \over
\Gamma(x)\Gamma(\beta-x)\Gamma(\beta+(n+j-1)k)\Gamma(k+1)}\,.
\end{equation}
Kaneko's identity gives directly the shifted hyperdeterminants as
\begin{eqnarray}
\fl D_{n;k}^{(k)}(r)
 =
{1\over n!}\left({1-\gamma\over\gamma}\right)^{nk(n-1)+nr}
\times\prod_{j=0}^{n-1}&
{\Gamma(x+jk+r)\Gamma(\beta-x+jk)\Gamma(\beta)\Gamma(jk+k+1) \over
\Gamma(x)\Gamma(\beta-x)\Gamma(\beta+r+(n+j-1)k)\Gamma(k+1)}\times
\nonumber\\ &\times P_{(n^r)}^{a,b}(p,\cdots,p;k)
\end{eqnarray}
with $a={x\over k}-1$, $b={\beta-x\over k}-1$ and $p={\gamma\over
\gamma-1}$.
\subsection{Krawtchouk}

The Krawtchouk polynomials are given by
\begin{equation}
K_n(x;p,N)={}_2F_1(-n,-x;-N; {1\over p})\,.
\end{equation}
Assuming at first that $-N$ is not a negative integer, we can
write down an integral representation
\begin{equation}\fl
K_n(x;p,N)= {\Gamma(-N)\over \Gamma(-x)\Gamma(-N+x)}\int_0^1
t^{-x-1}(1-t)^{-N+x-1}\left(1-{t\over p}\right)^ndt
\end{equation}
which leads immediately to the evaluation
\begin{equation}\fl
D_n^{(k)}(K)= {1\over n!}\left({\Gamma(-N)\over
\Gamma(-x)\Gamma(-N+x)}\right)^n \left(-{1\over
p}\right)^{kn(n-1)} S_n(-x,x-N,k)\,.
\end{equation}
After simplification, we find the expression
\begin{equation}
D_n^{(k)}(K)= {1\over n!p^{kn(n-1)}}
\prod_{j=0}^{n-1}{ (-x)_{jk} (N-x)_{jk} (jk+k)!\over (N)_{k(n+j-1)+r}k!}
\end{equation}
which is well defined for integral $N$, provided that all the
elements of the hyperdeterminant are also defined (recall that
$K_n$ is defined only for $n=0,\ldots,N$).

The shifted hypertur\'anians can be evaluated from Kaneko's
integral,
\begin{eqnarray}
\fl D_{n;r}^{(k)}(K)
 ={(-1)^{nr}\over n!k!^n}
\left({1\over p}\right)^{kn(n-1)+nr} \times\nonumber\\
\lo \qquad\quad\times
\prod_{j=0}^{n-1}
{
(-x)_{jk+r}(-N+x)_{jk}(kj+k)!\over
(-N)_{k(n+j-1)+r}}
P_{(n^r)}^{(\alpha,\beta)}(p,\ldots,p;k)
\end{eqnarray}
where $\alpha=-{x\over k}-1$ and $\beta={x-N\over k}-1$. In
particular, for $k=1$, we obtain a Wronskian of Jacobi polynomials
with parameters $-x-1$ and $-N+x-1$.

\section{Hankel hyperdeterminants and  symmetric functions}

In this section, we shall give an expression of the Hankel
hyperdeterminant $D_n^{(k)}(c)$ in terms of symmetric functions.
Precisely, we suppose here that $c_n=h_n(x)$, the $n$-th complete
homogeneous symmetric function of some auxiliary set of variables
$x=\{x_i\}$, and our aim is to obtain an expression of the
symmetric function $D_n^{(k)}(h)$ in terms of the Schur functions
$s_\lambda(x)$ (see \cite{Mcd} for  notation). It turns out
that this problem is equivalent to  finding the Schur expansion of
the even powers of the Vandermonde determinant, a difficult
problem which has been thoroughly discussed in recent years,
mainly in view of its potential applications to Laughlin's theory
of the fractional quantum Hall effect (see \cite{STW} and
references therein).

Since $D_n^{(k)}(h)$ is a homogeneous polynomial of degree $n$ in
the $h_i$, its Schur expansion will only involve partitions of
length at most $n$. We can therefore assume that
$x=\{x_1,\cdots,x_{n}\}$.

It will be convenient to work with Laurent polynomials in $x$. In
particular, for each vector
$\lambda=(\lambda_1,\cdots,\lambda_{n})\in\Z^n$, we define the
augmented monomial symmetric function
\[\tilde m_\lambda=\sum_{\sigma\in\SG_n}x^{\sigma\lambda}\]
where $\sigma(\lambda_1,\cdots,\lambda_n)=
(\lambda_{\sigma(1)},\cdots,\lambda_{\sigma(n)})$ and
$x^\lambda=x_1^{\lambda_1}\cdots x_n^{\lambda_n}$.

Let $\phi$ be the linear map sending $\tilde m_\lambda$ to
$h_\lambda$ if $\lambda\in\N^n$ and $0$ otherwise. As the set of
the $\tilde m_\lambda$, with $\lambda$ a decreasing sequence,
 is a basis of the space of symmetric Laurent
polynomials the map $\phi$ is well defined.

Let us consider now the alternants
\begin{equation}
a_\lambda=\sum_{\sigma\in\SG_n}\epsilon(\sigma)x^{\sigma\lambda}.
\end{equation}
The image by $\phi$ of the product of $2k$ alternants is a
hyperdeterminant. Since
\begin{equation}
\prod_{i=0}^{2k}a_{\lambda^{(i)}}=
\sum_{\sigma_1,\cdots,\sigma_{2k-1}\in \SG_n}\tilde
m_{\lambda^{(1)}+
\sigma_1\lambda^{(2)}+\cdots+\sigma_{2k-1}\lambda^{(2k)}}
\end{equation}
we get
\begin{equation}
\phi\left(\prod_{i=0}^{2k}
a_{\lambda^{(i)}}\right)=\DET_{2k}\left.\left(h_{\lambda_{i_1}^{(1)}
+\cdots+\lambda_{i_{2k}}^{(2k)} }\right)\right|_{1}^{n}.
\end{equation}
The case where $k=1$ is well know and can be found as an exercise
in the book \cite{Mcd}. It is shown there that for any symmetric
function $f$, we have
\begin{equation}
\phi(fa_\delta a_{-\delta})=f\,,
\end{equation}
where $\delta=(n-1,\ldots,2,1,0)$.  In particular,
\begin{eqnarray}\label{nvars}
D_n^{(k)}(h)&=\DET_{2k}(h_{i_1+\cdots+i_{2k}})_{0}^{n-1}=\phi(a^k_\delta)
=
\phi(\Delta(x)^{2k})\nonumber\\ 
&=
\phi\left((-1)^{n(n-1)\over2}\prod_{i=1}^{2k}
x_{i}^{n-1}\Delta(x)^{2(k-1)} a_{\delta}a_{-\delta}\right)\nonumber\\
&=(-1)^{n(n-1)\over2}\prod_{i=1}^{n}x_i^{n-1}\Delta(x)^{2(k-1)}\,.
\end{eqnarray}
Since we
are working with $n$ variables, the effect of the factor
$\prod_{i=1}^{n}x_i^{n-1}$ is to shift the parts of the partitions
occuring in the Schur expansion of $\Delta(x)^{2(k-1)}$ by $n-1$.

The Schur expansions of $\Delta^{2}$, which determines all Hankel
hyperdeterminants of order $4$, have been computed up to 9
variables in \cite{STW} and recently up to 10 variables by
Wybourne. Using the Littlewood-Richardson rule, it is
then possible to compute the powers $\Delta^{2k}$ for small values
of $k$. The first cases are
\begin{eqnarray*}
\fl D_2^{(2)}(h) =  -s_{31}+3s_{22}
\\ 
\fl D_3^{(2)}(h) = 
-s_{642}+3s_{633}+3s_{552}-6s_{543}+15s_{444}
\\ 
\fl D_2^{(3)}(h)  = -s_{51}+5s_{42}-10s_{33}
\\ 
\fl D_3^{(3)}(h) = -s_{10\,6\,2}+5s_{10\,5\,3}-10s_{10\,4\,4}
                 +5s_{972}-20s_{963}+25s_{954}
\\
\lo                -10s_{882}+25s_{873}+15s_{864}-100s_{855}
                  -100s_{774}+160s_{765}-280s_{666}\\
\fl D_2^{(4)}(h) =  -s_{71}+7s_{62}-21s_{53}+35s_{44}\,.
\end{eqnarray*}

In terms of the elementary symmetric functions $e_n$ and the power
sums $p_n$, this identity can be rewritten as
\begin{equation}\label{sym1}
D_n^{(k)}(h)=(-1)^{n(n-1)/2} e^{n-k}_n\det(p_{n-i+j})^{k-1}\,.
\end{equation}

As an illustration of (\ref{nvars}), let us consider the case
where $c_n=U_n(x)$, the $n$th Chebyshev polynomial of the second
kind. From the generating function
\[
\sum_{n\ge 0}U_n(x)t^n ={1\over 1-2xt+t^2}
\]
we see that $U_n(x)=h_n(x_1,x_2)$, where $x_1+x_2=2x$ and
$x_1x_2=1$. Hence, $\Delta^2=(x_1+x_2)^2-4x_1x_2=4(x^2-1)$, so
that $D_2^{(k)}(U)=-[4(x^2-1)]^{(k-1)}$. Comparing with
(\ref{apolar}), we obtain the identity
\begin{equation}
\frac12\sum_{j=0}^m(-1)^j{m\choose j}U_j(x)U_{m-j}(x)
=
\left\{
\begin{array}{cc} [4(1-x^2)]^{\frac m2-1} & \hbox{\rm $m$ even}\\
                   0 & \hbox{\rm otherwise}\,.
\end{array}\right.
\end{equation}
Specializing to the Fibonacci numbers $f_n=(-i)^nU_n(i/2)$,
(normalized such that $f_0=f_1=1$), we find
\begin{equation}
\frac12\sum_{j=0}^{2k}(-1)^j{2k\choose j}f_j f_{2k-j}=5^{k-1}\,.
\end{equation}

\subsection{An application: inverse factorials}

The symmetric function approach allows us to handle the case
\begin{equation}
c_n ={1\over n!}\,.
\end{equation}
Indeed, the image of a Schur function $s_\lambda$ under the
specialisation $h_n\mapsto c_n$ is equal to the scalar product
\begin{equation}
{1\over N!}\<s_\lambda, s_1^N\>
\end{equation}
where $N=|\lambda|$. If $\lambda$ has at most $n$ parts, we can
interpret the above as the scalar product of rational
$GL(n)$-characters, defined by
\begin{equation}
\<f(x),g(x)\>_{GL(n)} ={1\over n!}{\rm CT} \left\{
f(x)\prod_{i\not=j}\left(1-{x_i\over x_j}\right)g(\bar x) \right\}
\end{equation}
where CT means the constant term, and $\bar
x=(x_1^{-1},\ldots,x_n^{-1})$.

Hence, we can write
\begin{eqnarray}
\fl \If(k,n)= D_n^{(k)}(c) ={1\over [kn(n-1)]!} \left\<
(-1)^{n(n-1)\over 2}(x_1\cdots x_n)^{n-1}\Delta^{2k-2}(x)\,,\,
(x_1+\cdots +x_n)^{kn(n-1)} \right\> \nonumber\\
 \lo
={(-1)^{kn(n-1)\over 2}\over [kn(n-1)]!}{1\over n!}{\rm CT} \left\{
(x_1\cdots x_n)^{k(n-1)} \prod_{i\not=j}\left(1-{x_i\over
x_j}\right)^k \left(
 {1\over x_1}+\cdots +{1\over x_n}\right)^{kn(n-1)} \right\} \nonumber \\
\lo = {(-1)^{kn(n-1)\over 2}\over [kn(n-1)]!}
 \<(x_1\cdots x_n)^{k(n-1)}\,,\, (x_1+\cdots+x_n)^{kn(n-1)}\>'_\alpha
\end{eqnarray}
where in the last equation we have introduced Macdonald's second
scalar product $\<\ ,\ \>'_\alpha$ associated to Jack polynomials
in $n$ variables, with parameter $\alpha={1\over k}$ (see
\cite{Mcd}, (10.35) p. 183).

Now, $(x_1\cdots x_n)^{k(n-1)}=P_{(k(n-1))^n}^{(\alpha)}$, and
\begin{equation}
 (x_1+\cdots+x_n)^{kn(n-1)}
= \sum_{\kappa\vdash kn(n-1) } C_\kappa^{(\alpha)}(x)
\end{equation}
where, if $\kappa\vdash N$,
\begin{equation}
 C_\kappa^{(\alpha)}(x) = {\alpha^N N!\over c_\kappa(\alpha)}Q_\kappa^{(\alpha)}(x)
\end{equation}
with
\begin{equation}
c_\kappa(\alpha)=\prod_{(i,j)\in\kappa}(\alpha(\kappa_i-j)+\kappa'_j-i+1)\,.
\end{equation}

Therefore, if we denote by $\nu$ the rectangular partition
$(k(n-1))^n$ of weight $N=kn(n-1)$, the Hankel hyperdeterminant is
given by
\begin{equation}
\If(k,n)=
 {(-1)^{N}\over k^{N)} c_{\nu}(k^{-1}) }
\< P_\nu^{(1/k)}\,, Q_\nu^{(1/k)}\>'_{1/k}
\end{equation}
which can be evaluated in closed form thanks to equation (10.37)
of \cite{Mcd} p. 183. This yields
\begin{equation}
\fl \If(k,n)={(-1)^{kn(n-1)/2} (kn)!\over n!
(k!)^n}\prod_{i=0}^{n-1}{(ki)!\over(k(n+i-1))!}
\end{equation}
%
The case $k=1$ could have been obtained in a simpler
way from the hook-length formula giving the dimensions of the
irreducible representations of the symmetric group.

\section{Conclusion}

We have demonstrated that the calculation of Hankel hyperdeterminants
amounts to the evaluation of an interesting class of multidimensional
integrals, including Selberg's and Kaneko's ones, and more generally, the
partition functions of one-dimensional
Coulomb systems with logarithmic potential. We have presented a
series of examples which can be evaluated more or less directly
from known results, and obtained on our way a unified presentation
of many Hankel determinants, including some new cases.
However, obtaining new integrals from algebraic or combinatorial
evaluation of hyperdeterminants would be more interesting. A few
examples are presented in this paper, and we expect  more from
a systematic study of Hankel hyperdeterminants from an invariant theory
point of view. Indeed, any generalization  of one of 
the various tricks
working with ordinary Hankel determinants would immediately lead
to new interesting integrals.

\appendix

\section{Hyperdeterminantal aspects of Selberg's original proof}

The Selberg integral can be deduced from the Hilbert, factorial or
inverse factorial hyperdeterminants.  Actually, the evaluation of
any non-trivial class of Hankel hyperdeterminants would lead
either to Selberg's intergral in full generality, or to some
interesting generalization. This is already apparent in Selberg's
original proof. The first part of his proof can be translated in
terms of hyperdeterminants in the following way.

First, we write  Selberg's integral, for $c=k$ an integer, as a
hyperdeterminant
\begin{equation}
S_n(a,b,k)=n!B(\alpha,\beta)^nD^{(k)}_{n}(c(\alpha,\alpha+\beta))
\end{equation}
where $c_n(a,b)={(a)_n\over(b)_n}$ and $(a)_n=a(a+1)\cdots(a+n-1)$
is the Pochhammer symbol. Computing $S_n(a,b,k)$ amounts to find a
closed form for $D^{(k)}_{n}(c(a,b)).$

Expanding the hyperdeterminant, we find an expression of the type
\begin{eqnarray}
D^{(k)}_{n}(c(a,b))=&\sum_{J}c_{J}\prod_{i=1}^{n}{(a)_{j_l}\over(b)_{j_l}}
\\&=\prod_{m=1}^n{(a)_{k(m-1)}\over(b)_{k(n+m-2)}}
\sum_{J}c_{J}\prod_{m=1}^n{(a)_{j_m}(b)_{k(n+m-2)}\over
(a)_{k(m-1)}(b)_{j_m}}
\end{eqnarray}
where $J$ runs over  the integer vectors $J=(j_1,\cdots,j_n)$ such
that $j_1\leq j_2\leq \cdots j_n$,  $j_1+\cdots+j_n=kn(n-1)$, and
$c_{J}\in\Z$.

A straightforward investigation  of the $2k$-uplets of
permutations giving a non zero $c_{J}$ implies that for each
$m\in\{1,...,n\}$, one has
\begin{equation}
k(m-1)\leq j_m\leq k(n+m-2).
\end{equation}
Hence, our hyperdeterminant can be written as a product
\begin{equation}
D^{(k)}_{n}(c(a,b))=\prod_{m=1}^n{(a)_{k(m-1)}(b-a)_{k(m-1)}\over(b)_{k(n+m-2)}}
\times {P(a,b-a)\over Q(b-a)}
\end{equation}
where $P(a,b)$ is a polynomial whose degree in $b$ is at most
$\frac{kn(n-1)}2$ and $Q(b)$ is a polynomial of degree
$\frac{kn(n-1)}2$.

Since $D^{(k)}_{n}(c(a,b))$ is  symmetric in $a$ and $b-a$, we see
that the ratio ${P(a,b-a)\over Q(b-a)}=\alpha(n,k)$ is independent
of $a$ and $b$, so that
\begin{equation}
D^{(k)}_{n}(c(a,b))=\alpha(n,k)\prod_{m=1}^n
{(a)_{k(m-1)}(b-a)_{k(m-1)}\over(b)_{k(n+m-2)}}
\end{equation}
Now, setting $a=1$ and $b=2$ we obtain
\begin{equation}\fl
D^{(k)}_{n}(c(a,b))=\prod_{m=1}^n\frac{(1+k(m+n-2))!(a)_{k(m-1)}(b-a)_{k(m-1)}}{
(k(m-1))!^2(b)_{k(m+n-2)}}{\rm H}(k,n)
\end{equation}
that is, Selberg's integral can be deduced in full generality from
the Hilbert hyperdeterminant.

Also, observing that the case of inverse factorials is given by
the limit
\begin{equation}
\If(k,n)=\lim_{L\rightarrow\infty}
L^{-kn(n-1)}D^{(k)}_{n}(c(L+1,1))
\end{equation}
we have
\begin{equation}\fl
D^{(k)}_{n}(c(a,b))=(-1)^{kn(n-1)\over2}\prod_{m=1}^n
\prod_{m=1}^n{(a)_{k(m-1)}(b-a)_{k(m-1)}\over(k(n+m-2))!(b)_{k(n+m-2)}}\If(k,n)
\end{equation}
In the same way,  the Hankel hyperdeterminant of factorial numbers
\begin{equation}
{\rm
F}(n,k)=\lim_{L\rightarrow\infty}L^{kn(n-1)}D^{(k)}_{n}(c(1,L+1))
\end{equation}
gives another equivalent identity
\begin{equation}
D_n^{(k)}(c(a,b))=\prod_{m=1}^n{(a)_{k(m-1)}(b-a)_{k(m-1)}\over(k(m-1))!
(b)_{k(n+m-2)}}{\rm F}(n,k)\,.
\end{equation}

\section{Possible generalizations}

More generally, if we start with a hypergeometric moment sequence
$c_n={P(n)\over Q(n)}c_{n-1}$, where $P(n)=\sum_{i=0}^ra_in^i$ and
$Q(n)=\sum_{i=0}^sb_in^i$ are two polynomials in $n$, a similar
analysis leads to an expression of the hyperdeterminant
$D_{n}^{(k)}(c)$ as a product
\begin{equation}
D_n^{(k)}(c)=c_0^n\prod_{m=0}^{n-1}{\displaystyle\prod_{j=0}^{km-1}P(j)\over
\displaystyle\prod_{j=0}^{k(n+m-1)-1}Q(j)}R^{(k)}_n(\underline
a;\underline b)
\end{equation}
where $R^{(k)}_n(\underline a;\underline b)$ is a polynomial of
degree at most $kn(n-1)\over 2$ in  both sets of variables
$\underline a=\{a_0,\cdots,a_r\}$ and $\underline b=\{b_0,\cdots
b_s\}$ and whose coefficients are  in $\Z$.

In the most general case, $R^{(k)}_n(\underline a;\underline b)$
cannot  be factorized, and seems difficult to compute. However,
in some simple cases, we can give a closed form.

Suppose that $P(n)=cn^2+bn+a$ and $Q(n)=1$, we find
\begin{equation}
R^{(k)}_2(a,b,c;1)={(2k+1)!\over k!}\prod_{i=k+2}^{2k+1}(b+cj).
\end{equation}
Let us give now two examples involving combinatorial numbers.

The tri-Catalan numbers $C_n^{(3)}={\left(3n\atop n\right)\over
2n+1}$ admit a representation as moments \cite{PSo}
\begin{equation}
C_n^{(3)}=\frac{3^{3n+1}}{2\sqrt3\pi}\frac{B(n+\frac13,n+\frac23)}{2n+1}
=\int_0^{27\over
4}x^nd\mu(x)
\end{equation}
where
$d\mu(x)={\sqrt32^{\frac23}\over12\pi}
{2^{\frac13}(27+3\sqrt{81-12x})^{2\over3}-6x^{1\over3}\over
x^{2\over3}(27+3\sqrt{81-12x})^{1\over3}}dx$. 
It follows that our
hyperdeterminant has the integral representation
\begin{equation}
D_n^{(k)}(C^{(3)})=\frac1{n!}\int_0^{27\over4}\cdots\int_0^{27\over
4}\Delta(x)^{2k}\prod_{i=1}^nd\mu(x_i)
\end{equation}
which looks rather difficult to compute. Nevertheless, our
previous remarks allows to start the calculation
\begin{equation}
\fl
D_n^{(k)}(C^{(3)})=\prod_{m=0}^n{\prod_{j=0}^{km-1}(-2+11j-18j^2+9j^3)\over
\prod_{j=0}^{k(n+m-1)-1}(4j^2-j)}R_n^{(k)}(-2,11,-18,9;0,-1,4)
\end{equation}
and it remains to find a closed form for
$R_n^{(k)}(a_1,a_2,a_3,a_4;b_1,b_2,b_3)$. When $k=1$, the result
is known \cite{Tamm}.

If $c_n=(2n)!$ we have
\begin{eqnarray}
D_n^{(k)}(c)=&\frac1{2^n}\int_0^\infty\cdots\int_0^\infty
\Delta(x)^{2k}\prod_{i=1}^n{\exp(-\sqrt{x_i})\over\sqrt{x_i}}dx_i
\\&=2^{kn(n-1)}
\prod_{m=0}^n(mk-1)!\prod_{j=0}^{km-1}(2j-1)R_n^{(k)}(0,-2,4;0)
\end{eqnarray}

Let us give now some polynomials
$R_n^{(k)}(\underline a,\underline b)$ for various values of $n$,
$k$, $\underline a$ and $\underline b$:
\footnotesize
\begin{eqnarray*}
\fl R_2^{(1)}(a_0,a_1,a_2;b_0,b_1)=-a_0b_1 + a_1 b_0 + 3 a_2 b_0 + 2
a_2 b_1\\ 
\fl R_2^{(2)}(a_0,a_1,a_2;b_0,b_1)=6(-2 a_0a_1b_0b_1
     - 10 a_0a_2b_0b_1 + 15a_1a_2b_0b_1- a_0a_1b_1
     - 15 a_0a_2b_1^2  \\+ 9 a_1a_2b_0^2  + a_0^2b_1^2  + a_1^2b_0^2
   + 20 a_2^2b_0^2 + 24 a_2^2  b_1^2  + a_1^2  b_0 b_1 + 50 a_2^2  b_0 b_1)\\
\fl R_2^{(2)}(a_0,a_1,a_2,a_3;)=6(a_1^2 + 9 a_1 a_2 + 20 a_2^2  + a_0 a_3 + 35 a_1 a_3 + 150 a_2 a_3 + 274 a_3^2 )\\
\fl R_3^{(2)}(a_0,a_1,a_2;)=16(
    94251 a_0 a_1^3  a_2^2  + 5525 a_0^2  a_1^2  a_2^2
     + 48 a_0^3  a_1 a_2^2  + 1853066 a_0 a_1 a_2^4  \\ + 603101 a_0 a_1^2  a_2^3
         + 25518 a_0^2  a_1 a_2^3  + 7123 a_0 a_1^4  a_2 + 522 a_0^2  a_1^3  a_2
    + 3278390 a_1^3  a_2^3 \\ + 15303958 a_1^2  a_2^4  + 384 a_0^3  a_2^3
    + 41544 a_0^2  a_2^4  + 211 a_0 a_1^5  + 19 a_0^2  a_1^4  + 592 a_1^6
     \\+ 37115136 a_2^6  + 2178696 a_2^5  a_0 + 37277876 a_2^5  a_1 + 385834 a_1^4  a_2^2
       + 23654 a_1^5  a_2)\\
\fl R_2^{(2)}(;b_0,b_1,b_2,b_3)=-6(b_0 b_3 - b_1^2  - 11 b_1 b_2 - 45 b_1 b_3 - 30 b_2^2  - 250 b_2 b_3 - 524 b_3^2 )
\end{eqnarray*}
\normalsize

\section{Pseudo-hyperdeterminants}

For tensors of odd order,
another notion of hyperdeterminant is considered for example in  \cite{So1}
\begin{equation}
\DET_+(A_{i_1\cdots
i_{2k+1}})_0^{n-1}=\sum_{\sigma_1\cdots\sigma_{2k}}\epsilon(\sigma_1\cdots\sigma_{2k})
\prod_{i=1}^nA_{i\sigma_1(i)\cdots\sigma_{2k}(i)}.
\end{equation}
Note that this polynomial has not the same invariance properties
as the hyperdeterminant under  linear transformations.

When $A_{i_1 \cdots i_{2k+1}}=c_{m_{i_1}+i_2+\cdots+i_{2k+1}}$, the
$c_n$ being the moments of  a  mesure $d\mu(x)$, this
hyperdeterminant can be expressed a multiple integral involving an
even power of the Vandermonde determinant
\begin{eqnarray}
\fl _+D^{(k)}_{\underline
m}(c)=&\DET_{+}\left(c_{m_{i_1}+i_2+\cdots+i_{2k}}\right)_0^{n-1}\nonumber\\
&=\sum_{\sigma_1\cdots\sigma_{2k}}\epsilon(\sigma_1\cdots\sigma_{2k})\prod_{i=1}^{n}
\int_a^bx^{m_i+\sigma_1(i)+\cdots\sigma_{2k}(i)-2k}d\mu(x)\nonumber\\ 
&=
\int_a^b\cdots\int_a^b\sum_{\sigma_1\cdots\sigma_{2k}}\epsilon(\sigma_1\cdots\sigma_{2k})
\prod_{i=1}^nx_i^{m_i+\sigma_1(i)+\cdots\sigma_{2k}(i)-2k}d\mu(x_i)\nonumber\\
&=\int_a^b\cdots\int_a^b
\prod_{i=1}^nx_i^{m_i}\Delta(x)^{2k}d\mu(x_1)\cdots d\mu(x_n)
\end{eqnarray}
Obviously, one has
\begin{equation}
_+D^{(k)}_{(0^{n})}(c)={n!}D^{(k)}_{n}(c)
\end{equation}
and we can compute  other examples related to  Selberg's
integral. The main tool is the system of differential equations
verified by the functions $f_{\underline
m}=\Delta(x)^{2k}\prod_{i=1}^nx_i^{a+m_i-1}(1-x_i)^{b-1}$ (see
Aomoto's proof of a variant of the Selberg integral in \cite{Ao} or \cite{Meh1}
for example).

Let us list a few results:
\begin{enumerate}
\item For $c_n=\frac1{n+1}$, one has
\begin{eqnarray}
\fl  _+D^{(k)}_{(1^s,0^{n-s})}=&\int_0^1\cdots\int_0^1x_1\cdots
x_s\Delta(x)^{2k}dx_1\cdots dx_n\nonumber\\
&=\frac1{k!^n}\prod_{j=1}^s{1+(n-j)k\over
2+(2n-j-1)k}\prod_{j=0}^{n-1}{(k(1+j))!(kj)!^2\over(1+(n+j-1)k)!}
\end{eqnarray}
\item More generally, if $c_n={\Gamma(a+n)\over
\Gamma(b+n)}$, one gets
\begin{eqnarray}
\fl
_+D_{(1^s,0^{n-s})}^{(k)}(c)=&\frac{1}{\Gamma(b-a)^nk!^n}\prod_{j=1}^s{a+k(n-j)\over
b+(2n-j-1)k}\times\nonumber\\
&\times\prod_{j=0}^{n-1}{(k(1+j))!\Gamma(a+jk)\Gamma(b-a+jk)\over
\Gamma(b+(n+j-1)k)}
\end{eqnarray}
\item If $c_n=n!$, one obtains
\begin{equation}
\fl
_+D^{(k)}_{(1^s,0^{n-s})}(c)=\frac1{k!^n}\prod_{j=1}^m(1+k(n-j))\prod_{j=0}^{n-1}(k(1+j))!(kj)!
\end{equation}
and \begin{eqnarray}
\fl
_+D^{(k)}_{(2^m,1^s,0^{n-m-s})}(c)=&\frac1{k!^n}\prod_{j=1}^m(2+k(2n-m-j))\times
\nonumber\\
&\times\prod_{j=1}^{m+s}(1+k(n+j))\prod_{j=0}^{n-1}(k(1+j))!(kj)!
\end{eqnarray}
\end{enumerate}

\section*{References}

\end{document}